\documentclass[a4paper,12pt,final]{article}
\usepackage{etex}

\usepackage[sc]{mathpazo}

\usepackage{setspace,graphicx,amsmath,amsfonts,amssymb,amsthm}
\usepackage{marginnote,enumitem,subfigure,rotating,fancyvrb}
\usepackage{float}
\usepackage[longnamesfirst]{natbib}

\usepackage[dvipsnames]{xcolor}

\usepackage{morefloats}     
\usepackage{subfigure}     
\usepackage{ctable}
\usepackage{colortbl}
\usepackage{longtable}
\usepackage{rotating}
\usepackage[english]{babel}
\usepackage{amsmath} 
\usepackage{amssymb}           
\usepackage[all,color]{xy}
\usepackage{ragged2e}
\usepackage{ifpdf}
\usepackage{color}
\usepackage{parskip}
\usepackage{marvosym}

\usepackage{titlesec}
\newcommand{\periodafter}[1]{#1.}
\titleformat{\subsubsection}[runin]
{\normalfont\bfseries}{\thesubsubsection}{1em}{\periodafter}

\usepackage{anyfontsize}
\usepackage{t1enc}

\usepackage{caption}
\DeclareCaptionFont{captiofont}{\fontsize{9pt}{10.8pt}\selectfont}
\usepackage[lmargin=1in, rmargin=1in, tmargin=1.5in, bmargin=1.5in, paperwidth=220mm, paperheight=290mm]{geometry}

\makeatletter\let\chapter\@undefined\makeatother 

\usepackage[plainpages=false,
breaklinks=true,
colorlinks=true,
urlcolor=gray,
citecolor=blue,
linkcolor=blue,
bookmarks=true,
bookmarksopen=true,
bookmarksopenlevel=1,
pdfstartview=FitH,
pdfview=FitH]{hyperref}

\usepackage{color,hyperref}

\usepackage[symbol]{footmisc}

\def\x{\mbox{\boldmath$x$}}

\def\bbeta{\mbox{\boldmath$\beta$}}

\def\balpha{\mbox{\boldmath$\alpha$}}

\begin{document}

\renewcommand{\thefootnote}{\arabic{footnote}}
\setcounter{footnote}{0}

\setlist{noitemsep}  

\newcommand{\tit}{\vspace{1em}\Large Deep Learning, Predictability, and Optimal \\Portfolio Returns \footnote{Jozef Barunik gratefully acknowledges support from the Czech Science Foundation under the 25-18070S project.}}

\newcommand{\abs}{\small 
We study the dynamic portfolio selection of an investor who uses deep learning methods to forecast stock market excess returns. In a two-asset allocation problem, deep neural networks -- both feedforward and long short-term memory (LSTM) recurrent architectures -- deliver economically significant gains in terms of certainty equivalent returns and Sharpe ratios relative to linear predictive regressions. These gains are robust to alternative performance measures, the inclusion of transaction costs, borrowing and short-selling constraints, different rebalancing horizons, and subsample splits, and are particularly pronounced during NBER recessions and periods with large return swings. Within the class of neural networks we consider, economic performance is broadly similar across architectures, with the recurrent LSTM specification providing incremental benefits with more frequent rebalancing. Overall, our evidence suggests that exploiting the time-series structure of standard predictor variables via deep learning can generate meaningful portfolio improvements for investors beyond those obtained from linear models.\vspace{0.3in}
\\
\textbf{Keywords:} Return predictability, portfolio allocation, machine learning, recurrent neural networks, empirical asset pricing
\\
\\
\textbf{JEL codes:} C45, C53, E37, G11, G17}

\title{\vspace{-2em}{\bf\tit}}

 \author{Mykola Babiak\setcounter{footnote}{6}\thanks{Department of Accounting and Finance, Lancaster University Management School, Lancaster, LA1 4YX, UK. E-mail: \url{m.babiak@lancaster.ac.uk }\,\, Web: \href{https://sites.google.com/site/mykolababiak/home}{sites.google.com/site/mykolababiak}} \hspace{2em} \and \hspace{2em} Jozef Barun\'{i}k\setcounter{footnote}{11}\thanks{Institute of Economic Studies, Charles University, Opletalova 26, 110 00, Prague, CR and Institute of Information Theory and Automation, Academy of Sciences of the Czech Republic, Pod Vodarenskou Vezi 4, 18200, Prague, Czech Republic. E-mail: \url{barunik@utia.cas.cz }\,\, Web: \href{https://barunik.github.io/}{barunik.github.io}}}

\date{\hspace{2em}}

\maketitle
\thispagestyle{empty}

\vspace*{0.5in}
\centerline{\bf Abstract}
\medskip
\abs
\normalsize
\setlength{\parskip}{.2cm }
\setlength{\parindent}{0.6cm}

\doublespacing

\clearpage
\pagenumbering{arabic}


\section{Introduction}

An extensive empirical asset pricing literature has documented support for the predictability of equity returns.\footnote{See, for example, \cite{campbell1987stock}, \cite{campbell1988dividend}, \cite{fama1988dividend}, \cite{fama1989business}, \cite{ferson1991}, \cite{pesaran1995predictability}, \cite{lettau2001consumption}, \cite{lewellen2004predicting} and \cite{ang2007stock}, among many others. } With an ever-increasing number of potential predictors, the practice of applying machine learning methods to make the most accurate predictions using large datasets is gaining further traction. This emerging literature demonstrates the superior performance of machine learning relative to linear regression analysis, which is favoured by researchers for its simplicity.\footnote{\cite{welch2007comprehensive} uses around 20 financial and macroeconomic variables to predict aggregate market returns. \cite{green2013supraview} lists more than 330 return prediction signals used by the existing literature over the period 1970-2010. \cite{harvey2016and} report 316 ``factors'' useful for predicting stock returns.} However, it is unclear whether the solid statistical performance of machine learning translates into portfolio gains for an investor who uses these models of return predictability to construct optimal portfolios.\footnote{The existing evidence on the use of linear models suggests that an ensemble of additional features is needed to improve portfolio performance derived from linear predictive regressions. Additional ingredients include learning about predictability with informative priors \citep{wachter2009predictable} and an ensemble of estimation risk and time-varying volatility \citep*{johannes2014sequential}.} Furthermore, this literature typically does not work with the number of time series properties essential for the timing of stock returns.

In this paper, we contribute to the existing evidence by exploring the value of predictions offered by deep neural networks, including a long short-term memory (LSTM) recurrent architecture that excels at learning nonlinear time-series dependencies for portfolio construction. In contrast to the extensive literature on the statistical and economic significance of standard feedforward neural networks, which exploit the information contained in the cross-section of returns, we introduce the LSTM recurrent structure due to its various advantages. LSTM can capture a complex nonlinear time series dependence structure, including long-run dependence and nonstationarity, and it mitigates the problem of gradient vanishing and gradient exploding to some extent by adding gating units to capture the dependence. We argue that time series dependence is an important feature to explore, and we demonstrate that LSTM recurrent neural networks can effectively identify time series dependence and deliver context-specific portfolio gains.

Specifically, we demonstrate the economic value of deep machine learning methods in a canonical asset pricing problem: predicting the aggregate US stock market index.\footnote{For the literature on market return prediction, see \cite{welch2007comprehensive}, \cite{koijen2010predictability}, \cite{Rapach:2010}, \cite{cochrane2011presidential}, \cite{rapach2021asset}, among others.} Our rationale for applying deep LSTM recurrent NNs to the prediction of a single stock market index is motivated by the virtue of complexity in the US equity market prediction \citep{kelly2022virtue}. To evaluate the LSTM recurrent network, we also implement a variety of machine learning architectures, including shallow and deep feedforward neural networks. For ease of comparison with the existing literature, the inputs to our models are the standard predictor variables considered by \cite{welch2007comprehensive}. We then examine the asset allocation of an investor with a power utility, who chooses between a market portfolio and a risk-free asset, using forecasts of deep NNs.\footnote{The single asset prediction boils down to a time-series problem of market timing and can be extended to the setting with many assets. In the latter case, the role of asset return covariances becomes important. We focus on the market return alone to isolate the time-series dimension of return prediction, while exploiting the cross-sectional predictability of asset returns would certainly improve our main results.} We compare the results with benchmark portfolios that exploit market predictability via linear models. As the empirical analysis is based on the standard dataset, our results are comparable to portfolios using (penalised) linear models \citep{johannes2014sequential,rapach2021asset} and tree-based models \citep{rossi2018predicting}. For this reason, we do not implement the portfolios based on other methods and focus mainly on deep NNs in our empirical investigation.

We find that deep neural networks deliver economically significant gains for long-term investors. By departing from the expectations hypothesis and exploiting return predictability through our best-performing LSTM specification, the investor’s annual certainty-equivalent return roughly doubles (from around 4.7\% to approximately 10\%), while the monthly Sharpe ratio increases from 0.049 to approximately 0.175. These gains are robust to alternative measures of portfolio performance (such as cumulative return, maximum drawdown, and maximum one-month loss) and to the inclusion of transaction costs, short selling, and borrowing constraints. Similar conclusions are obtained when a timing strategy is implemented.

We then attempt to better understand the properties of the portfolio gains provided by neural networks. For quarterly rebalancing, the LSTM portfolio typically generates particularly high CERs and SRs during NBER recessions, while all deep networks yield substantial improvements over linear benchmarks during both recessions and expansions. Over the postwar sample, the LSTM architecture attains the highest CERs and SRs in five of the seven decades. However, simpler feedforward networks occasionally match or exceed its performance, particularly during the expansions of the 1990s and 2010s when the US economy did not experience a major recession. Investors also benefit more from using LSTM when they rebalance their portfolios more frequently. With annual rebalancing, however, the incremental gains from the recurrent architecture are substantially weaker.

When we compare the various neural network specifications, we observe that increasing architectural complexity does not necessarily improve portfolio performance, even though all deep learning models achieve a similar level of statistical accuracy. In our study, deep feedforward networks and recurrent LSTMs demonstrate broadly comparable economic performance, with LSTMs providing modest, context-dependent enhancements, especially during adverse periods, rather than achieving uniform dominance. \cite{nagel2025seemingly} recently showed that a strategy based on a complex nonlinear transformation of a small number of features simplifies to a volatility-timed momentum strategy. This helps to explain why machine-learning portfolios can outperform those based on linear models. \cite{cujean2017does} also demonstrate that time-series momentum intensifies during recessions. Taken together with our findings, these results imply that the advantages of deep learning primarily stem from the models' capacity to leverage persistent state variables and time-varying dynamics. Furthermore, they suggest that explicitly capturing time-series dependence provides additional benefits beyond incorporating nonlinearities into predictive regressions.

Machine learning has a long history in economics and finance \citep*{hutchinson1994nonparametric,kuan1994artificial,racine2001nonlinear,baillie2007testing}. At its core, one may perceive machine learning as a general statistical analysis that economists can use to capture hidden, complex relationships when using simple linear methods. \cite*{breiman2001statistical} emphasize that maximizing prediction accuracy in the face of an unknown model differentiates machine learning from the more traditional statistical objective of estimating a model assuming a data-generating process. Using innovative optimization techniques, machine learning seeks to choose the most preferred model from an unknown pool of models. However, problems in finance differ from typical machine learning applications in the importance of time-series dependence, a key property we aim to focus on in this paper.

Machine learning applications in finance have begun to emerge \citep*{heaton2017deep,messmer2017deep,bryzgalova2019forest,bianchi2020bond,gu2020empirical,tobek2020does,zhang2020deep,freyberger2020dissecting,chen2024deep}. These studies employ deep learning methods to improve the cross-sectional prediction of individual stock returns.\footnote{Additional applications of machine learning methods in mortgage and credit markets include \cite{sirignano2016deep} and \cite{fuster2018predictably}.} Our key contribution to this literature is to explore the portfolio gains when predicting stock market returns via deep learning, where the time-series effects become more prominent. 

Our research builds on the work of \cite{rapach2021asset}, who examined the predictability of aggregate stock market returns using an extension of the conventional predictive regression approach. In contrast, we implement non-linear machine learning models. In this respect, our paper is most closely related to the work of \cite{feng2018deep}, who evaluated the statistical performance of deep neural networks in predicting stock market returns, and \cite*{chen2020deep}, who modelled a time-varying stochastic discount factor using recurrent architectures. Our research question examines market return predictability and a two-asset portfolio problem using the \cite{welch2007comprehensive} dataset. It is also related to non-linear tree-based models in \cite{rossi2018predicting} and penalised linear models in \cite{rapach2021asset}. 

Relative to this literature, our contribution is threefold. First, we apply deep neural networks (both feedforward and recurrent) to a classic low-dimensional market-timing setting based on Welch–Goyal predictors, explicitly studying the dynamic portfolio choice of a representative investor. Second, rather than focusing exclusively on statistical forecast metrics or SDF fit, we evaluate models using utility-based portfolio criteria (certainty-equivalent returns, Sharpe ratios, and timing strategies) across rebalancing horizons and business-cycle states. Third, we relate network-based importance diagnostics to non-parametric measures of serial dependence in the predictors. This provides new evidence on the time-series properties and state variables that drive the economic value of deep learning in this environment. In a recent publication, \cite*{jiang2020re} demonstrate how convolutional neural networks (a class of deep learning model) can be used to learn price patterns from images in a very different setting. The substantial portfolio gains of complex networks documented in our paper are consistent with the results of \cite{kelly2022virtue}.

The remainder of this paper is organized as follows. Section \ref{section: Evaluating Predictability via Portfolio Performance} discusses standard approaches to assessing expected return predictability, introduces nonlinear machine learning methods, outlines details of hyperparameter tuning, describes the portfolio choice problem of an investor, and outlines performance measures. Section \ref{section: Empirical Results} describes the data and summarizes the main results. Section \ref{section: Further Analysis} assesses the variable importance in driving optimal portfolio performance, performs the subsample analysis, identifies economic drivers of portfolio performance, provides robustness checks to alternative performance measures, transaction costs, borrowing and short-selling constraints, different rolling window sizes, and finally implements a timing strategy. Section \ref{section: Conclusion} concludes.

\section{Evaluating Predictability via Portfolio Performance}
\label{section: Evaluating Predictability via Portfolio Performance}

The standard approach used to forecast excess equity returns is a linear model of the form
\begin{equation}
r_{t+1} = \alpha + \beta x_t + \varepsilon_{t+1}^{r},
\label{eq: linear regression}
\end{equation}
where 
$r_{t+1}$
are monthly log excess returns, 
$\alpha$
and
$\beta$
are coefficients to be estimated, 
$x_t = (x_t^1, ..., x_t^n)$
is a set of predictor variables, and
$\varepsilon_{t+1}^{r}$
is a normal error term. A large strand of empirical literature has examined linear regression models with multiple predictors including prominent variables such as the dividend yield, valuation ratios, various interest rates and spreads, among others.\footnote{See, for example, \cite*{shiller1981stock}, \cite*{hodrick1992dividend}, \cite*{stambaugh1999predictive}, \cite*{avramov2002stock}, \cite*{cremers2002stock}, \cite*{ferson2003spurious}, \cite*{lewellen2004predicting}, \cite*{torous2004predicting}, \cite*{campbell2006efficient}, \cite*{ang2007stock}, \cite*{campbell2008predicting}, \cite*{cochrane2008dog}, \cite*{lettau2008reconciling}, and \cite*{pastor2009predictive}.} Although researchers have proposed numerous variables for predicting stock market returns, empirical evidence on the degree of predictability is mixed at best. \cite*{welch2007comprehensive} find that most linear specifications with multiple predictors perform poorly and remain insignificant even in-sample. They further show that an investor using linear models to forecast equity returns would not be able to improve portfolio performance compared to a no predictability benchmark. 

There are several reasons for the lack of robust evidence on the predictability of stock market returns via linear models. The specification defined by Equation (\ref{eq: linear regression}) assumes a linear and time-invariant relationship between log excess returns and predictors, which is at odds with the theoretical and empirical evidence.\footnote{Leading examples include \cite*{menzly2004understanding}, \cite*{paye2006instability}, \cite*{santos2006labor}, \cite*{lettau2008reconciling}, and \cite*{henkel2011time,Dangl:2012}.} Bayesian learning about uncertain parameters in the linear regression has been proposed as a way to introduce a time-varying relationship between the returns and predictor variables. However, sequential parameter learning leads to significant portfolio benefits only in the presence of a highly informative prior \citep*{wachter2009predictable} or a combination of estimation risk and time-varying volatility \citep*{johannes2014sequential}. Thus, prior knowledge about the nature of expected return predictability or careful modeling of its conditional features, especially time variation in return volatility, are critical for generating economic gains.

This paper takes an alternative approach inspired by recent developments in machine learning in the empirical asset pricing literature.\footnote{Leading studies include \cite*{giglio2017inference}, \cite*{kelly2015three}, \cite*{heaton2017deep}, \cite*{feng2018deep}, \cite*{feng2019deep}, \cite*{chen2024deep}, \cite*{kelly2019characteristics}, \cite*{freyberger2020dissecting}, \cite*{gu2020empirical} and \cite*{kozak2020shrinking}. } Specifically, we use recurrent neural network structures to approximate the functional relationship between the set of predictors and returns for optimal portfolio construction. We do not impose a known form of this relationship but instead allow for the flexible identification of potentially nonlinear time-series interactions from the data. Our choice of LSTM recurrent neural networks over other machine learning methods (e.g., tree-based approaches) is motivated by the fact that they offer ways to explore important time-series dependencies in the data. This paper aims to revisit the evidence documented in \cite*{welch2007comprehensive} and to show that, unlike linear predictive regression, the sound statistical performance of neural networks specialized in time series translates into substantial portfolio improvements for an investor using these novel methods when dynamically constructing an optimal portfolio.

A machine learning approach to equity return prediction involves selecting the most preferred model from an unknown pool of models using innovative optimization techniques. More formally, we can define our prediction problem by characterizing excess equity returns $r_{t+1}$ using a set of predictor variables $x_t$ as:
\begin{equation}
r_{t+1} = \mathfrak{f}_{W,b}(x_t) + \varepsilon_{t+1}^{r},
\end{equation}
where $\mathfrak{f}_{W,b}$ is a model to be found by neural network with weight matrices $W=\left(W^{(1)},\ldots,W^{(L)}\right)$ and bias vector $b=\left(b^{(1)},\ldots,b^{(L)}\right)$ in $L$ layers such as
\begin{equation}
\widehat{r}_{t+1} := \mathfrak{f}_{W,b}(x_t) = f^{(L)}_{W^{(L)},b^{(L)}} \circ \ldots \circ f^{(1)}_{W^{(1)},b^{(1)}} \left(x_t\right),
\label{eq: ffnet}
\end{equation}
with $f^{(\ell)}_{W^{(\ell)},b^{(\ell)}} := f_{\ell}\left( W^{(\ell)}x_t + b^{(\ell)} \right) = f_{\ell}\left( \sum_{i=1}^{m} W^{(\ell)}_i x_t + b_i^{(\ell)} \right)$ being commonly chosen as sigmoidal (e.g. $f_{\ell}(z) = 1/(1+\exp(-z))$), or rectified linear units (ReLU) ($f_{\ell}(z) = \max\{z,0\}$). Note that in case functions $\mathfrak{f}$ are linear, $\mathfrak{f}_{W,b}(x_t)$ is a simple linear regression, regardless of the number of layers $L$. For example with $L=2$, the model becomes a reparametrized simple linear regression: $\widehat{r}_{t+1} = W^{(2)} (W^{(1)}x_t + b^{(1)}) + b^{(2)}=\beta x_t + \alpha,$ and hence the model is generalization of Equation (\ref{eq: linear regression}). In case $\mathfrak{f}_{W,b}(x_t)$ is nonlinear, neural network complexity grows with increasing $m$, and with increasing the number of hidden layers $L$, or growing deepness of the network, we have a deep neural network.

\begin{figure}[t!]
\caption{(Deep) Recurrent Network}
\vspace{10pt}
\begin{justify}
\footnotesize\centering This figure illustrates a deep recurrent neural network model.
\end{justify}
\vspace{10pt}
\centering\includegraphics[width=\textwidth]{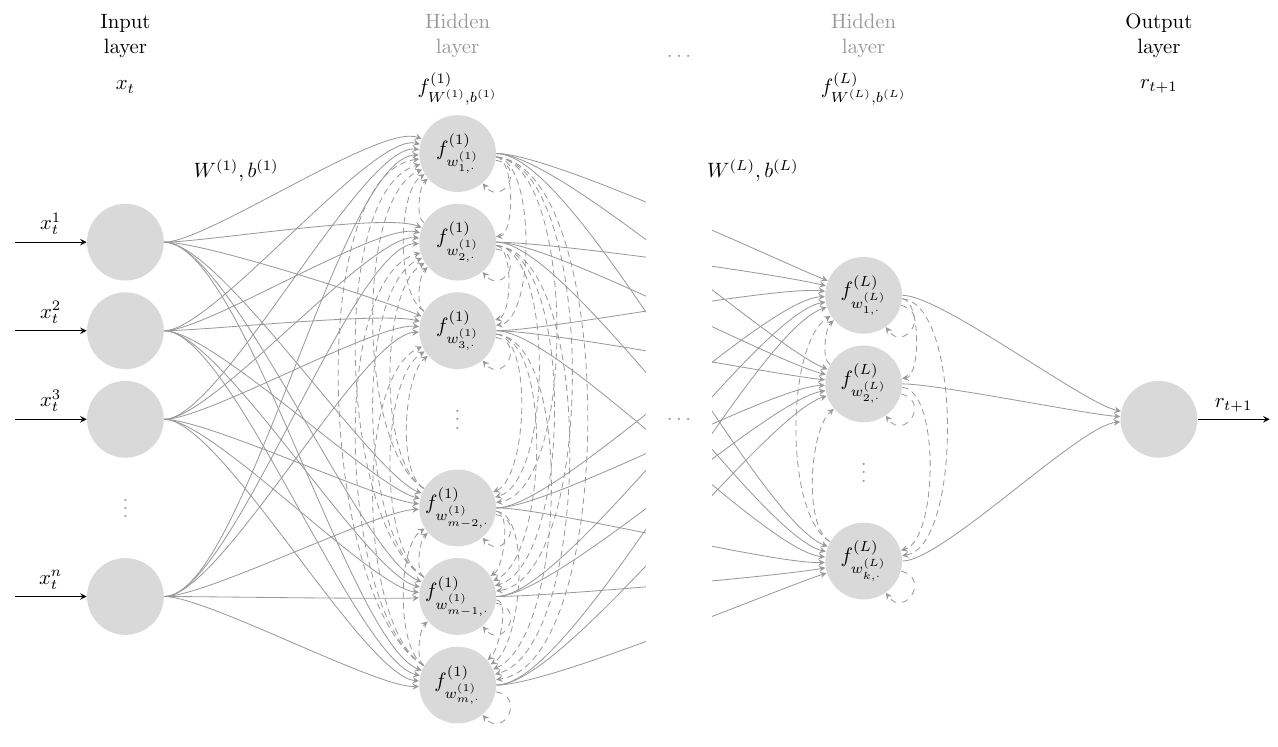}
\label{fig: RNnet}
\end{figure}

\textbf{LSTM (Deep) Recurrent Networks.} While many predictor variables used in finance are not i.i.d. and evolve dynamically over time, a recurrent neural network (RNN) that considers time series behaviour can help in the prediction task. The LSTM is designed to find hidden state processes, allowing for lags of unknown and potentially long-term dynamics in the time series. Specifically, RNN  introduces a time series dependence into Equation (\ref{eq: ffnet}) by transforming a sequence of predictors into another sequence via a lagged hidden state:
\begin{equation}
h_{t} = f(W_h h_{t-1} + W_x x_t + b_0).
\label{eq: RRnet1}
\end{equation}
Intuitively, RNN is a nonlinear generalization of an autoregressive process where lagged variables are transformations of the lagged observed variables. Figure \ref{fig: RNnet} depicts $W_h$ using dashed lines and $W_x$ using solid lines and illustrates how the network structure additionally uses lagged information. In Equation (\ref{eq: RRnet1}), this structure is only useful when the immediate past is relevant. In case the time-series dynamics are driven by events that are further back in the past, the addition of complex LSTMs that incorporate memory units \citep*{hochreiter1997long} is required.

Memory units allow the network to learn when to forget previous hidden states and when to update hidden states given new information. Specifically, in addition to a hidden state, LSTM includes an input gate, a forget gate, an input modulation gate, and a memory cell. The memory cell unit combines the previous memory cell unit, which is modulated by the forget and input modulation gates together with the previous hidden state, modulated by the input gate. These additional cells enable an LSTM to learn highly complex long-term temporal dynamics that a vanilla RNN is not capable of. Such structures can be viewed as a flexible hidden state space model for a large-dimensional system. Additional depth can be added to the LSTM by stacking them on top of each other, using the 
hidden state of the LSTM as the input to the next layer.

In Equation (\ref{eq: ffnet}), a new memory cell $c_t$ is created with the current value of a predictor variable $x_t$ and the previous hidden state $h_{t-1}$, and is then combined with a forget gate that controls the amount of information stored in the hidden state as
\begin{eqnarray}
h_{t} &=& \sigma\left(\underbrace{W_h^{(o)} h_{t-1} + W_x^{(o)} x_t + b_0^{(o)}}_{\text{output gate}}\right) \circ \tanh (c_t) \\
c_{t} &=& \sigma\left(\underbrace{W_h^{(g)} h_{t-1} + W_x^{(g)} x_t + b_0^{(g)}}_{\text{forget gate}}\right) \circ c_{t-1} + 
\sigma\left(\underbrace{W_h^{(i)} h_{t-1} + W_x^{(i)} + b_0^{(i)}}_{\text{input gate}} \right) \circ \tanh(k_t).
\label{eq: RRnet2}
\end{eqnarray}
The term $\sigma(\cdot)\circ c_{t-1}$ introduces the long-range dependence, and $k_t$ is the new information flow to the current cell. The states of forget and input gates control the weights of memory and new information. In Figure \ref{fig: RNnet}, $c_t$ is the memory pass through multiple hidden states in the recurrent network.

\subsection{Estimation, Hyperparameters, Details}

Due to the high dimensionality and nonlinearity of the problem, estimation of a deep neural network is a complex task. Here, we provide a detailed summary of the model architectures and their estimations. We work with a variety of deep learning structures and compare them with a recurrent LSTM network and regularized OLS. We consider NN1, NN2, and NN3 models that contain 16, 32--16, and 32--16--8 neurons in the one, two, and three hidden layer structures, respectively, and an LSTM model, which is an NN with three recurrent layers with 32-16-8 neurons in each and LSTM cells introduced into the last layer. 

To prevent the model from over-fitting and reduce many parameters, we use dropout, a common form of regularization that generally performs better than traditional $l_1$ or $l_2$ regularization. The term dropout refers to dropping out units in neural networks and can be shown to be a form of ridge regularization. To fit the networks, we adopt a popular and robust adaptive moment estimation algorithm (Adam) with weight decay regularization introduced by \cite*{kingma2014adam} and we use the Huber loss function in the estimation.

Further, we follow the most common approach in the literature and select tuning parameters adaptively from the data in a validation sample. We perform this hyperparameter tuning each period the investor rebalances the portfolio. The investor is assumed to rebalance her portfolios with a quarterly (annual) frequency. Therefore, we need to find the new optimal hyperparameters quarterly (annually), which alleviates computational costs. At the end of each quarter (year), we split the data into training and validation samples that maintain the temporal ordering of the data and tune hyperparameters with respect to the statistical and economic criteria. We search the optimal models in the following grid of 100 randomly chosen combinations of the following hyperparameters: learning rate $\in[0.001,0.02]$, decay regularization $\in[0,0.001]$, dropout $\in[0\%,60\%]$ of weights and activation function $\in\{\text{sigmoid},\text{ReLU}\}$ with 1000 epochs with early stopping. Since the sample at each window is rather small and final models can depend on initial values in the optimization, we use ensemble averaging of five models with randomly chosen initial values.\footnote{We have estimated our models on two servers with 48 core Intel\textregistered~ Xeon\textregistered~ Gold 6126 CPU@ 2.60GHz and 24 core Intel\textregistered~ Xeon\textregistered~ CPU E5-2643 v4 @ 3.40GHz, 768GB memory and two NVIDIA  GeForce RTX 2080 Ti GPUs. We have used \texttt{Flux.jl} with \textsf{JULIA} 1.4.0. for the model fitting. A complete rolling window estimation with hyperparameter tuning takes around two days. We have confirmed that our estimation results are robust to using a larger hyperparameter space. As a full hyperparameter search on a larger hyperparameter space can easily take weeks or months, even on our fast GPU cluster, we have selectively tested further hyperparameters.}

\subsection{Optimal Portfolios}
\label{section: optimal portfolios}
We consider a portfolio choice problem of an agent with an investment horizon of 
$T$
periods in the future who maximizes her expected utility over the cumulative portfolio return. There are two assets: a one-period Treasury bill and a stock index.\footnote{Extending our analysis to multiple assets is straightforward; however, we consider a portfolio choice problem with two assets as in \cite*{barberis2000investing} and more recently \cite*{johannes2014sequential} and \cite*{rossi2018predicting} to make our results directly comparable to other studies.} If
$\omega_{t+\tau}$
is the allocation to the stock index at time
$t+\tau,$ 
the investor solves the following optimization problem at time 
$t$
\begin{equation}
\label{eq: utility maximization}
\max_{\omega} \mathbb{E}_t\left[U(r_{p,t+T})\right]
\end{equation}
in which the end-of-horizon portfolio return 
$r_{p,t+T}$
is defined as
\begin{equation}
\label{eq: cumulative return}
r_{p,t+T} = \prod_{\tau=1}^{T} \left[(1-\omega_{t+\tau-1})\exp(r_{t+\tau}^f) + \omega_{t+\tau-1}\exp(r_{t+\tau}^f + r_{t+\tau})\right],
\end{equation}
and 
$r_{t+\tau}^f$ denotes a zero-coupon default-free log bond yield between 
$t+\tau-1$
and
$t+\tau$.
Following \cite*{johannes2014sequential}, we consider various choices of horizons 
$T$
to assess the impact of the length of the investment period, that is, we allow the investor to rebalance portfolio weights with different frequencies. Specifically, we report the results for the two cases of six months
$(T=6)$
and two years
$(T=24).$
Although these choices of the rebalancing and investment horizons correspond to the modeling approach of \cite*{johannes2014sequential}, one can also implement a monthly rebalancing strategy. We consider less frequent trading to alleviate computational costs related to recursive hyperparameter tuning of neural networks. 

The allocations between a Treasury bill and a stock index are updated every three months, or once per year for the shorter or longer investment horizons, respectively. These choices of horizons and rebalancing periods allow us to compare two investment strategies. The former reflects a more actively managed portfolio with frequent changes in the allocations, whereas the latter corresponds to a relatively passive investment portfolio with less frequent rebalancing. We further winsorize the weights for the stock index to 
$-1 \leq \omega_{t+\tau} \leq 2$
to prevent extreme investments.\footnote{\cite{johannes2014sequential} restrict portfolio weights to satisfy 
$-2 \leq \omega_{t+\tau} \leq 3$
and indicate that the portfolio returns from the OLS models perform much worse with wider intervals or uncapped weights. We bound portfolio weights to a narrower interval to make the OLS models more stable while allowing short-selling.} In the sensitivity analysis, we check the robustness of our results to alternative assumptions about the portfolio weights, particularly incorporating the borrowing and short-selling constraints.

We also assume a power utility investor
\begin{equation*}
U(r_{p,t+\tau}) = \frac{r_{p,t+\tau}^{1 - \gamma}}{1-\gamma},
\end{equation*}
where 
$\gamma$
is the coefficient of risk aversion. The expected utility is defined by the predictive distribution of cumulative portfolio returns
$r_{p,t+\tau}$ 
given by Equation (\ref{eq: cumulative return}), which in turn depends on the corresponding model used to predict future excess returns 
$r_{t+\tau}$
and the law of motion of predictor variables
$\x_t.$ 
For 
$\x_t,$ 
we adopt a parsimonious AR(1) framework, that is, each variable 
$\x_t^i$
satisfies 
\begin{equation*}
\x_t^i = \balpha^{x^i} + \bbeta^{x^i}\x_{t-1}^i + \varepsilon_{t}^{x^i}.
\end{equation*}
where 
$\balpha^{x^i}$
and
$\bbeta^{x^i}$
are coefficients, and
$\varepsilon_{t}^{x^i}$
are normal error terms. To proxy for the joint variance-covariance matrix  of the error terms 
$\varepsilon_{t} = (\varepsilon_{t}^r,\varepsilon_{t}^x),$ 
we employ a sample variance estimator 
$\hat{\Sigma}_t = \hat{\varepsilon}_t \hat{\varepsilon}_t^{'},$
where 
$\varepsilon_t$
are forecast errors. Finally, we set the risk aversion parameter
$\gamma = 4$
to compare our results to the existing literature \citep*{johannes2014sequential,rossi2018predicting}.

In sum, the investor maximizes expected utility and optimally rebalances portfolio weights quarterly or annually for six-month and two-year investment horizons. To compute the expected utility and optimal weights, the agent uses the distribution of returns predicted by the linear regressions or the neural networks.

\subsection{Numerical Solution for Optimal Portfolio Weights}

The numerical solution for optimal portfolio weights closely follows the procedure outlined in \cite{barberis2000investing} and \cite{johannes2014sequential}. At time 
$t,$ 
the investor estimates a particular predictive model $\mathcal{M}_s$ for stock market returns and AR(1) processes for individual predictors using the historical data. Given the recent observations of the risk-free rate 
$r_{t}^f,$ 
predictors 
$x_{t}^{i},$
and the model prediction
$\mathfrak{f}^{\mathcal{M}_s},$ the investor then aims to maximize the expected utility
\begin{equation}
\label{eq: utility maximization appendix}
\max_{\omega} \mathbb{E}_t\left[U(r_{p,t+T})\right],
\end{equation}
subject to constraints
\begin{equation}
\label{eq: cumulative return appendix}
r_{p,t+T} = \prod_{\tau=1}^{T} \left[(1-\omega_{t+\tau-1})\exp(r_{t+\tau}^f) + \omega_{t+\tau-1}\exp(r_{t+\tau}^f + r_{t+\tau})\right],
\end{equation}
\begin{equation}
\label{eq: stock market return and risk-free rate}
r_{t+\tau} = \mathfrak{f}^{\mathcal{M}_s}(\x_{t+\tau-1}) + \varepsilon_{t+\tau}^{r} \quad \wedge \quad r_{t+\tau}^f = r_{t}^f 
\end{equation}
\begin{equation}
\label{eq: x equations}
\x_{t+\tau}^i = \balpha^{x^i} + \bbeta^{x^i}\x_{t+\tau-1}^i + \varepsilon_{t+\tau}^{x^i},
\end{equation}
\begin{equation}
\label{eq: errors}
\varepsilon_{t+\tau} = \left(\varepsilon_{t+\tau}^r,\varepsilon_{t+\tau}^x\right) \sim \mathcal{N}(0,\hat{\Sigma}_t),  \quad \tau = 1,...,T
\end{equation}
where the variance-covariance matrix of the error terms 
$\hat{\Sigma}_t = \hat{\varepsilon}_t \hat{\varepsilon}_t^{'}$
is a sample estimator based on the residuals in the estimation window. Note that we make several assumptions to reduce the computational costs in computing the expected utility. First, we assume a constant risk-free rate $r_{t+\tau}^f = r_{t}^f$ in Equation (\ref{eq: stock market return and risk-free rate}), which alleviates the need for its modeling. Second, we adopt a parsimonious AR(1) framework given by Equation (\ref{eq: x equations}) for each predictive variable, which makes the computation of future values $\x_{t+\tau}^i$ straightforward. Third, we assume that the variance-covariance matrix in Equation (\ref{eq: errors}) remains the same in future periods, again simplifying the computations. 

We calculate the integral in Equation (\ref{eq: utility maximization appendix}) for a two-dimensional grid of $\omega = (\omega_1,\omega_2)$ and choose 
$\omega$ that maximizes expected utility.\footnote{Note that there are two weights $\omega_1$ and $\omega_2$ due to quarterly (annual) rebalancing with a six-month (two-year) horizon.} For the main results, we restrict the allocation to the interval 
$-1 \leq \omega \leq 2,$ whereas we consider alternative restrictions in robustness checks. We evaluate the integrals by simulation. Specifically, we generate 1,000,000 independent draws of 
$\varepsilon_{t+\tau}$ 
from the Normal distribution 
$\mathcal{N}(0,\hat{\Sigma}_t),$ then iterate Equations (\ref{eq: cumulative return appendix})-(\ref{eq: x equations}) forward with the model prediction 
$\mathfrak{f}^{\mathcal{M}_s}$
and the parameters 
$\left(\balpha^{x^i}, \bbeta^{x^i}\right)$ obtained at time $t$ to compute portfolio returns 
$r_{p,t+T}$, and finally average 
$U(r_{p,t+T})$
over all draws.

\subsection{The Choice of Predictive Models}

To evaluate the impact of the investor's conditioning information, we consider different assumptions about the set of predictors and sample periods used to estimate the models. In particular, we consider the following specifications:
\begin{itemize}
\item[1.] The expectations hypothesis (EH) framework assumes a constant mean and constant variance framework with no predictors in Equation (\ref{eq: linear regression}), that is, $\bbeta = 0.$
\item[2.] A simple linear regression of excess log returns with the dividend yield as a single predictor and a ``kitchen sink'' linear regression with all available variables. We further implement OLS regressions for each case using all data up to time $t$ or over a 10-year rolling window, as in \cite*{johannes2014sequential}. The univariate models with the expanding and rolling windows are denoted OLS1 and OLS2, and the multivariate versions are OLS3 and OLS4.
\item[3.] A set of machine learning architectures including neural networks with 1 layer of 16 neurons (NN1), 2 layers of 32-16 neurons (NN2), 3 layers of 32-16-8 neurons (NN3), and the LSTM model with 3 recurrent layers and 32-16-8 neurons and LSTM cells introduced in the last layer. All NNs use a ``kitchen sink'' approach by utilizing all available data to predict log excess returns and are trained on a 10-year rolling window to account for a time-varying relationship between the predictors and returns. 
\end{itemize}

There are many dimensions that can be used to generalize our modeling approach. More general specifications could add additional predictor variables \citep*{mccracken2016fred}, parameter uncertainty \citep*{wachter2009predictable,johannes2014sequential,bianchi2020sparse}, economic restrictions \citep*{van2010predictive}, or consider a larger set of investable assets and alternative preferences \citep*{dangl_weissensteiner_2020}, among other extensions. Most notably, modeling stochastic volatility via a parsimonious mean-reverting process \citep*{johannes2014sequential} or more complex GARCH- and MIDAS-type volatility estimators \citep*{rossi2018predicting} would certainly improve the performance of our strategies. Instead, we consider all specifications with a constant volatility setting to solely evaluate the impact of neural networks on the performance of dynamic allocation strategies. We aim to demonstrate out-of-sample portfolio gains from using deep learning in the most restrictive setting. 

\subsection{Performance Evaluation}

In our analysis, we employ several metrics measuring the statistical accuracy of the methods considered and their economic gains for the investor. Concerning the statistical performance, we first compute the mean squared prediction error (MSPE) defined as
\begin{equation}
MSPE_q = \frac{1}{N_q}\sum_{t \in \mathcal{T}_q}\Big(r_t - \hat{r}_t^{\mathcal{M}_s}\Big)^2, \quad
MSPE_a = \frac{1}{N_a}\sum_{t \in \mathcal{T}_a}\Big(r_t - \hat{r}_t^{\mathcal{M}_s}\Big)^2,
\end{equation}
where
$r_t$
denotes the excess log return, 
$\hat{r}_t^{\mathcal{M}_s}$
is the return predicted by a model
$\mathcal{M}_s,$
$\mathcal{T}_q$
and
$\mathcal{T}_a$
are end-of-quarter or end-of-year months when we construct the model predictions and form the optimal portfolios,
$N_q$
and
$N_a$
denote the number of observations. Notice that the investor rebalances the portfolio at varying frequencies. Thus, we compute the prediction errors only in those periods when she reoptimizes her portfolio.

As in \cite*{campbell2008predicting}, we compute the out-of-sample $R^2$ for quarterly and annual rebalancing:
\begin{equation}
R^2_{q,oos} = 1 - \frac{\sum\limits_{t \in \mathcal{T}_q}\Big(r_t - \hat{r}_t^{\mathcal{M}_s}\Big)^2}{\sum\limits_{t \in \mathcal{T}_q}\Big(r_t - \bar{r}_t\Big)^2}, \quad
R^2_{a,oos} = 1 - \frac{\sum\limits_{t \in \mathcal{T}_a}\Big(r_t - \hat{r}_t^{\mathcal{M}_s}\Big)^2}{\sum\limits_{t \in \mathcal{T}_a}\Big(r_t - \bar{r}_t\Big)^2},
\label{equatin: r2oos}
\end{equation}
where 
$\bar{r}_t$
is the historical mean of returns. By construction, the out-of-sample $R^2$ statistic compares the out-of-sample performance of the chosen model 
$\mathcal{M}_s$
relative to the historical average forecast. Notice that we compute the historical mean over the same sample used to estimate 
$\mathcal{M}_s,$
which corresponds to either an expanding sample or a 10-year rolling window. The positive value of 
$R^2_{oos}$
indicates that the model-implied forecast has a smaller mean squared predictive error compared to the error implied by the historical average forecast. Thus, we perform a formal test of the null hypothesis
$R^2_{oos}\leq0$
against the alternative hypothesis
$R^2_{oos}>0$
by implementing the MSPE-adjusted \cite{Clark:West:2007} test. Note that we calculate the \cite{Clark:West:2007} test only if 
$R^2_{oos}$
is positive.

After we compare different models in terms of the statistical accuracy of their predictions, we assess whether a superior statistical fit translates into economic gains. It is worth noting that this relationship is non-trivial. Indeed, \cite*{campbell2008predicting} and \cite*{Rapach:2010} note that seemingly small improvements in 
$R_{oos}^2$
could generate large benefits in practice. We start our investigation of the size of the improvements by calculating the average Sharpe ratio of portfolio returns as a common measure of portfolio performance used in finance. The drawback of this metric is that it does not take tail behaviour into account. Consequently, we follow \cite{fleming2001economic} and compute the certainty equivalent return (CER) by equating the utility from CER to the average utility implied by an alternative model. Finally, we visualize the performance of all specifications by plotting the cumulative log portfolio returns over the sample period considered. This allows us to clearly see the time intervals in which the investor benefits the most from using different frameworks.

To evaluate the statistical significance of portfolio gains, we follow \cite{bianchi2020bond} and implement the test \'{a} la \cite{diebold1995comparing}. Specifically, we perform a pairwise comparison between the CERs generated by each framework under consideration and those yielded by the EH specification.\footnote{For the significance of SRs, we first need to simulate artificial returns under a null model of no predictability, that is, a model with constant mean and constant volatility. For each simulation, we need to obtain the forecasts for all models considered and construct optimal portfolios. Since a complete exercise of hyperparameter tuning takes around 2 days on the supercomputer cluster, repeating it, say, 500 times will increase cluster computing time proportionally. This makes the task computationally infeasible given the current computing capacity, unless more resources for parallel computing become available.} For each model 
$\mathcal{M}_s,$
we estimate the regression
\begin{equation*}
\mathcal{U}_{t+T}^{\mathcal{M}_s} - \mathcal{U}_{t+T}^{EH} = \alpha^{\mathcal{M}_s} + \varepsilon_{t+T}, 
\end{equation*}
where 
$\mathcal{U}_{t+T}^{X} = \frac{\left(r_{p,t+T}^{X}\right)^{1-\gamma}}{1-\gamma}$
and
$r_{p,t+T}^{X}$
is the cumulative portfolio return with the horizon 
$T.$
Testing for the difference in the CERs boils down to a test for the significance in 
$\alpha^{\mathcal{M}_s}.$

\section{Empirical Results}
\label{section: Empirical Results}

\subsection{Data and preliminary results}

Our empirical analysis of the S\&P 500 excess return predictability is based on the applications of various linear models and nonlinear machine learning methods as discussed in Section \ref{section: optimal portfolios}. We use a set of economic predictor variables considered by \cite{welch2007comprehensive} to compare our results to the literature. Specifically, we focus on the monthly historical data of twelve predictors, including dividend yield (dy), log earning price ratio (ep), dividend payout ratio (de), book-to-market ratio (bm), net equity expansion (ntis), treasury bill rates (tbl), term spread (tms), default yield spread (dfy), default return spread (dfr), cross-sectional premium (csp), inflation growth (infl), and stock variance (svar).\footnote{The data are retrieved from Amit Goyal's website and are available via the following \href{https://docs.google.com/spreadsheets/d/1bM7vCWd3WOt95Sf9qjLPZjoiafgF_8EG/edit?usp=sharing&ouid=113571510202500088860&rtpof=true&sd=true}{\textbf{link}} as of 27th May 2024.}$^{,}$\footnote{Compared to other studies on equity premium prediction, \cite{welch2007comprehensive} employ several other predictors (for example, investment-to-capital and consumption-to-wealth ratios, among others), which are available only for quarterly and annual frequencies. \cite{rossi2018predicting} includes twelve predictors similar to our study. \cite{kelly2022virtue} use fifteen variables by adding one lag of the market return, dividend-to-price ratio (dp), and long-term yield (lty). We do not consider the lagged market return similar to \cite{welch2007comprehensive} and omit dp and lty due to the high correlation with dy and tbl.}

Table \ref{tab:statistical accuracy} reports the statistical accuracy of the models considered. Panels A and B show the MSPE and out-of-sample
$R^2$
statistics based on those periods when quarterly and annual rebalancing happens. As shown in Panel A, all linear regressions yield larger MSPEs than the constant-mean, constant-volatility model, whereas neural networks best fit the data. 

\begin{table}[t!]
\centering
\footnotesize
\caption{Statistical Accuracy of Excess Return Forecasts}
\begin{minipage}{\textwidth} 
This table reports the mean-squared prediction error and out-of-sample 
 $R^2_{oos}$
obtained from using different methodologies to predict future S\&P 500 excess returns as outlined in Section \ref{section: optimal portfolios}. We compute the out-of-sample
 $R^2_{oos}$
in comparison to the expectations hypothesis using the historical mean to predict returns. Panel A shows the results when the investor maximizes a 6-month portfolio return and changes the allocations quarterly. Panel B demonstrates the results for a 2-year horizon and annual rebalancing. We compute statistical accuracy measures in those periods when the investor reevaluates her allocations with quarterly or annual frequency. We also report a p-value (in parentheses) of the null hypothesis 
 $R^2_{oos}\leq0$
following \cite{Clark:West:2007}. We report statistical significance only if 
 $R^2_{oos}$
is positive. The forecast starts in February 1955. The sample period spans from January 1945 to December 2018.
\end{minipage}
\vspace{10pt}

\begin{tabular}{lccccccccc}
\toprule
& \multicolumn{1}{c}{EH} & \multicolumn{1}{c}{OLS1} & \multicolumn{1}{c}{OLS2} & \multicolumn{1}{c}{OLS3} & \multicolumn{1}{c}{OLS4} & \multicolumn{1}{c}{NN1} & \multicolumn{1}{c}{NN2} & \multicolumn{1}{c}{NN3} & \multicolumn{1}{c}{LSTM} \\
\midrule
\multicolumn{10}{l}{Panel A: 6-month horizon and quarterly rebalancing}      \\
\midrule
$MSPE_q\times10^4$  & \cellcolor{YellowGreen!10}17.4  & \cellcolor{red!20}18.0    & \cellcolor{red!20}18.2  & \cellcolor{red!20}18.9  & \cellcolor{red!20}18.0    & \cellcolor{YellowGreen!20}16.3  & \cellcolor{YellowGreen!20}16.6  & \cellcolor{YellowGreen!20}16.6  & \cellcolor{YellowGreen!10}17.3 \\

$R^2_{q,oos}$ & \cellcolor{YellowGreen!10}0.0\% & \cellcolor{red!20}-2.5\% & \cellcolor{red!20}-3.6\% & \cellcolor{red!30}-8.0\%  & \cellcolor{red!20}-2.5\% & \cellcolor{YellowGreen!20}7.1\% & \cellcolor{YellowGreen!25}5.1\% & \cellcolor{YellowGreen!30}5.6\% & \cellcolor{YellowGreen!10}1.6\% \\
p-value & & & & & & (0.006) & (0.007) & (0.002) & (0.002) \\
\midrule
\multicolumn{10}{l}{Panel B: 2-year horizon and annual rebalancing}      \\
\midrule
$MSPE_a\times10^4$  & \cellcolor{Goldenrod!15}14.2  & \cellcolor{Orange!30} 15.0    & \cellcolor{Orange!30}14.7  & \cellcolor{Orange!40}16.7  & \cellcolor{Orange!50}19.9  & \cellcolor{Goldenrod!40}12.2  & \cellcolor{Goldenrod!60}11.1  & \cellcolor{Goldenrod!40}11.9    & \cellcolor{Goldenrod!40}12.1 \\

$R^2_{a,oos}$ & \cellcolor{Goldenrod!10}0.0\% & \cellcolor{Orange!20}-2.8\% & \cellcolor{Orange!20}-0.8\% & \cellcolor{Orange!40}-15.0\% & \cellcolor{Orange!50}-36.9\% & \cellcolor{Goldenrod!40}16.0\%  & \cellcolor{Goldenrod!65}23.8\% & \cellcolor{Goldenrod!50}17.6\% & \cellcolor{Goldenrod!50}17.2\% \\
p-value & & & & & & (0.001) & (0.014) & (0.007) & (0.008) \\
\bottomrule
\end{tabular}%
\label{tab:statistical accuracy}%
\end{table}%

A multivariate linear regression does not necessarily outperform a univariate model. Indeed, a linear regression estimated using a rolling window (OLS3) yields a noisier result and a larger MSPE than regressions using only the dividend yield (OLS1), whereas the ``kitchen sink'' linear regression with an expanding-window estimation (OLS4) slightly outperforms a single-predictor model (OLS2). Furthermore, in line with the findings of \cite{welch2007comprehensive}, none of the linear regressions outperform the simple historical mean, as indicated by the negative out-of-sample $R^2$. By contrast, all of the deep learning methods achieve a positive $R^2$ value, indicating the statistical benefits of accounting for non-linear relationships between stock market returns and predictors, as demonstrated by \cite{feng2018deep} and \cite{rossi2018predicting}. A formal test confirms that the expected return predictability generated by neural networks differs statistically from a naive historical mean forecast. Unreported results show that, within the class of neural networks considered here, differences in MSPE and $R^2$ are small and not statistically significant, meaning that the LSTM does not outperform the best feedforward architectures purely statistically.

Panel B also shows the results favoring NNs in a setting with less frequent rebalancing. In this case, the out-of-sample 
$R^2$
statistics increase substantially to around $20\%,$ which seems to be quite high at first glance. The reason is that 
$R^2_{a,oos}$ defined by Equation (\ref{equatin: r2oos}) computes the prediction errors in months when the models are re-estimated and portfolios are rebalanced (that is, each twelfth month). Thus, the absolute values of our statistics are not directly comparable to those reported in \cite{feng2018deep} who predict the stock market returns each month. This also applies to $R^2_{q,oos}$ based on observations (actual returns and predictions) at the end of each quarter. To have a better sense of why these statistics are amplified (especially for the annual rebalancing), we compute the autocorrelation of the end-of-quarter  (corresponding to the quarterly rebalancing) and end-of-year (corresponding to the annual rebalancing) monthly stock market returns. The autocorrelation is -0.02 in the former case and increases to 0.12 in the latter case. Thus, the substantial persistence in the time series contributes to the increased magnitude in
$R^2_{a,oos}.$
Finally, \cite{feng2018deep} report the out-of-sample $R^2$ of around $13\%$ when predicting the 12-month stock market returns with deep neural networks and a cumulative window, which can also be explained by the strong persistence in annual returns.

\subsection{Portfolio Results}

Table \ref{tab: portfolio returns cer and sr} provides a summary of annualized CERs and monthly SRs of portfolio returns for each model assuming a 6-month (Panel A) and 2-year (Panel B) rebalancing period. The summary statistics in each panel are computed for the whole sample and for recessions and expansions as defined by the NBER recession indicator. 

We recover a standard result for traditional methods: linear regressions do not generate out-of-sample improvements as measured by the CERs compared to the constant mean and constant volatility model. Regarding model-generated SRs, linear models perform slightly better than the expectations hypothesis model, with higher Sharpe ratios for more predictor variables. The rolling-window estimation introduces time-varying slope coefficients and leads to modest improvements. However, ignoring the estimation risk and stochastic volatility of returns results in lower CERs relative to a constant mean and volatility specification. 

The magnitude of performance measures is generally consistent with \cite{johannes2014sequential}. For instance, the certainty equivalent returns of two strategies based on the historical mean and volatility of market returns of 4.7\% and 4.5\% correspond well to the range of values from 4.5\% to 5.0\% reported in \cite{johannes2014sequential}. Further, smaller and even negative certainty equivalent returns from linear models align with the common wisdom documented by prior literature. The monthly Sharpe ratios of around 0.05 documented in our paper are slightly below Sharpe ratios ranging from 0.08 to 0.10 in \cite{johannes2014sequential}. This discrepancy in the results is attributable to differences in the sample periods. Our empirical analysis focuses on the postwar period when the stock market performance was slightly weaker as measured by lower Sharpe ratios, while \cite{johannes2014sequential} perform their empirical analysis on the data from 1927.

Turning to neural networks, we observe that the improved $R_{oos}^2$ achieved with machine learning methods directly translates into economic gains for an investor. Across all architectures, deep networks deliver substantially higher certainty equivalent returns and Sharpe ratios than the expectations hypothesis benchmark. For instance, with quarterly rebalancing in Panel A of Table \ref{tab: portfolio returns cer and sr}, our primary LSTM model produces an annual CER of approximately 10\% and a monthly SR of 0.175, compared to 4.7\% and 0.049 for the historical-mean model. Other deep feedforward networks achieve CERs and Sharpe ratios close to these values. Therefore, the main distinction lies between nonlinear neural network models and linear predictive regressions, rather than between specific neural network designs.

In general, when we compare NN1 to NN3 and the LSTM, we find that increasing the complexity of the architecture does not necessarily improve portfolio performance, even though all neural networks are statistically similar in terms of forecast accuracy. The LSTM is a three-layer network, making it directly comparable to NN3 in terms of structural complexity. It often performs exceptionally well for quarterly rebalancing and during recessions, though it is occasionally matched or slightly outperformed by simpler feedforward networks, particularly at the annual rebalancing horizon. A one-sided test confirms that the portfolio performance of neural networks is significantly better than that generated by the expectations hypothesis model, with the exception of NN3. Furthermore, comparing Panels A and B demonstrates that investors benefit more from using deep networks when managing their portfolios actively. Overall, these results suggest that the predictability of expected returns, as captured by nonlinear methods, provides valuable information for portfolio construction. The recurrent architecture offers incremental gains in specific environments rather than providing a uniformly superior solution.

These portfolio improvements are generally of the same magnitude as portfolio gains investors could generate by applying advanced linear models or other nonlinear predictive methods, such as tree-based models. For instance, \cite{johannes2014sequential} demonstrate that linear specifications with stochastic volatility can produce certainty equivalent returns ranging from 5\% to 8\% on an annual basis and Sharpe ratios ranging from 0.12 to 0.17 on a monthly basis. Most of the deep NNs considered in our empirical analysis produce performance within these ranges, with the best-performing LSTM generating an annualized CER of around 10\% and a monthly Sharpe ratio of 0.175. 

Comparing our results with other nonlinear methods reported in the literature, \cite{rossi2018predicting} document that investment portfolios based on boosted regression trees generate Sharpe ratios in the range from 0.12 to 0.16 depending on various approaches for modeling conditional volatility of stock returns. The author also shows that this performance is stronger than the Sharpe ratios from linear models, which generally belong to the interval from 0.03 to 0.08 on a monthly basis. The numbers reported in Table \ref{tab: portfolio returns cer and sr} are comparable with those in \cite{rossi2018predicting}. Further, it seems that deep NNs can provide a further boost in portfolio performance without explicitly modeling the conditional stock return volatility.

\begin{table}[t!]
\centering
\footnotesize
\caption{Certainty Equivalent Returns and Sharpe Ratios}
\begin{minipage}{\textwidth} 
This table reports the annualized certainty equivalent returns and monthly Sharpe ratios for different models outlined in Section \ref{section: optimal portfolios}. Panel A shows the results when the investor maximizes a 6-month portfolio return and changes the allocations quarterly. Panel B shows the results for a 2-year horizon and annual rebalancing. Each panel computes the statistics for the whole sample, with expansion and recession periods as defined by NBER. For the statistical significance of CERs, we report a one-sided p-value (in parentheses) of the test \'{a} la \cite{diebold1995comparing}. In particular, we regress the difference in utilities for each model
$\mathcal{M}_s$
and EH:
$\mathcal{U}_{t+T}^{\mathcal{M}_s} - \mathcal{U}_{t+T}^{EH} = \alpha^{\mathcal{M}_s} + \varepsilon_{t+T},$
where 
$\mathcal{U}_{t+T}^{X} = \frac{\left(r_{p,t+T}^{X}\right)^{1-\gamma}}{1-\gamma}$
and
$r_{p,t+T}^{X}$
is the cumulative portfolio return with the horizon 
$T.$
Testing for the difference in the CERs boils down to a test for the significance in 
$\alpha^{\mathcal{M}_s}.$
We flag in \textbf{bold font} CER values that are significant at the 10\% confidence level. The portfolio construction starts in February 1955. The sample period spans from January 1945 to December 2018.
\end{minipage}
\vspace{10pt}

\begin{tabular}{lccccccccc}
\toprule
& \multicolumn{1}{c}{EH} & \multicolumn{1}{c}{OLS1} & \multicolumn{1}{c}{OLS2} & \multicolumn{1}{c}{OLS3} & \multicolumn{1}{c}{OLS4} & \multicolumn{1}{c}{NN1} & \multicolumn{1}{c}{NN2} & \multicolumn{1}{c}{NN3} & \multicolumn{1}{c}{LSTM} \\
\midrule
\multicolumn{10}{l}{Panel A: 6-month horizon and quarterly rebalancing}      \\
\midrule
1955-2018 &       &       &       &       &       &       &       &       &  \\
[0.5em]

CER   & 4.737 & 2.643 & -0.030 & 2.781 & 2.491 & \textbf{7.295} & \textbf{6.984} & 5.491 & \textbf{10.007} \\
p-value &       & (1.000) & (1.000) & (0.935) & (0.954) & (0.027) & (0.032) & (0.292) & (0.000) \\
SR    & 0.049 & 0.046 & 0.062 & 0.088 & 0.095 & 0.166 & 0.157 & 0.144 & 0.175 \\
\midrule
Expansions &       &       &       &       &       &       &       &       &  \\
[0.5em]

CER    & 4.948 & 3.073 & -0.173 & 4.598 & 2.045 & 5.873 & 5.280  & 5.304 & \textbf{7.998} \\
p-value &       & (0.998) & (1.000) & (0.654) & (0.982) & (0.258) & (0.398) & (0.403) & (0.000) \\
SR    & 0.100   & 0.077 & 0.048 & 0.092 & 0.108 & 0.149 & 0.143 & 0.135 & 0.149 \\
\midrule
Recessions &       &       &       &       &       &       &       &       &  \\
[0.5em]

CER   & 3.311 & -0.274 & 1.401 & -9.079 & 6.752 & \textbf{19.024} & \textbf{20.806} & 6.936 & \textbf{26.770} \\
p-value &       & (0.995) & (0.648) & (0.944) & (0.221) & (0.000) & (0.000) & (0.200) & (0.000) \\
SR    & -0.193 & -0.182 & 0.154 & 0.091 & 0.036 & 0.284 & 0.255 & 0.204 & 0.358 \\
\midrule
\multicolumn{10}{l}{Panel B: 2-year horizon and annual rebalancing}      \\
\midrule
1955-2018 &       &       &       &       &       &       &       &       &  \\
[0.5em]

CER   & 4.542 & 1.068 & 0.040  & 0.923 & -0.067 & \textbf{6.342} & \textbf{6.879} & \textbf{6.437} & \textbf{5.622} \\
p-value &       & (1.000) & (1.000) & (0.999) & (0.997) & (0.000) & (0.000) & (0.000) & (0.012) \\
SR    & 0.048 & 0.044 & 0.046 & 0.083 & 0.081 & 0.138 & 0.136 & 0.129 & 0.118 \\
\midrule
Expansions &       &       &       &       &       &       &       &       &  \\
[0.5em]

CER   & 4.448 & 0.826 & 0.514 & 0.321 & 0.231 & \textbf{6.051} & \textbf{6.390} & \textbf{5.913} & \textbf{5.537} \\
p-value &       & (1.000) & (1.000) & (0.999) & (0.987) & (0.002) & (0.000) & (0.000) & (0.011) \\
SR    & 0.100   & 0.076 & 0.037 & 0.097 & 0.137 & 0.136 & 0.149 & 0.132 & 0.112 \\
\midrule
Recessions &       &       &       &       &       &       &       &       &  \\
[0.5em]

CER   & 5.235 & 2.866 & -2.924 & 5.930  & -2.138 & \textbf{8.611} & \textbf{10.975} & \textbf{10.836} & 6.353 \\
p-value &       & (0.997) & (1.000) & (0.262) & (1.000) & (0.006) & (0.000) & (0.000) & (0.284) \\
SR    & -0.190 & -0.170 & 0.102 & 0.017 & -0.104 & 0.160  & 0.111 & 0.149 & 0.154 \\
\bottomrule
\end{tabular}%
\label{tab: portfolio returns cer and sr}%
\end{table}%

We dissect this superior performance by looking at portfolio return statistics in periods of expansion and recession. Panel A in Table \ref{tab: portfolio returns cer and sr} shows that economic gains generated by NNs are large during both regimes and are especially pronounced in recessions. For instance, the annualized CER generated by the LSTM is, on average, around 8\% in good times, which is more than the 5\% predicted by the EH model. In bad times, the difference in performance is extremely large, with around 26\% and 3\% CERs in the LSTM and EH models, respectively. A pairwise test confirms that the improvement of LSTM over EH is statistically significant during expansions and recessions. In contrast, the portfolio returns of NN1 through NN3 are indistinguishable from EH in expansions, while shallower networks exhibit significantly better performance in recessions. In terms of the Sharpe ratios, the investor who ignores expected return predictability experiences, on average, around -19\% Sharpe ratios in recessions. In contrast, the LSTM model helps generate significant portfolio gains around 36\% SRs, with other NNs generating at least 20\% SRs on a monthly basis. Further, all NNs outperform linear regressions across good and bad times. 

These results are consistent with the existing literature showing that stock market predictability is concentrated in bad times \citep{rapach2016short,rapach2021asset}.\footnote{\cite{gargano2019bond} report a similar pattern for bond returns. Recently, \cite{bianchi2020bond} show that bond return predictability is also present in expansions when machine learning methods are employed.} There could be several reasons why we observe this typical pattern across all neural network designs. First, \cite{Rapach:2010} demonstrate that combining forecasts incorporates information from numerous economic variables and reduces forecast volatility, leading to improved out-of-sample performance, particularly during recessions. Note that the final prediction from neural networks is an ensemble of forecasts. Thus, we argue that both plain vanilla and recurrent NNs likely outperform linear models due to model averaging, which reduces forecast volatility. Second, \cite{cujean2017does} show that time-series momentum strengthens in recessions. Thus, we should expect improved results for models that could better identify this persistence in bad times. Consistent with this conjecture, LSTM produces a small increment in performance metrics compared to NN1-NN3 because it is designed to better capture time series dependence. Third, \cite{Dangl:2012} point out the importance of time-varying coefficients. Thus, time-varying hyperparameters might play an additional role in our results. 

Recently, \cite{nagel2025seemingly} examines the economic explanation for the strong performance of a stock market index timing
strategy based on the prediction of Random Fourier Features (RFF), a complex nonlinear transformation of a small number of features in a small sample. The author shows that the RFF-based strategy boils down to a volatility-timed momentum strategy. We conjecture that the outperformance of the NN-based portfolios could also be driven by a similar mechanism. Furthermore, the stronger results for the LSTM model compared to other NNs could be explained by the ability of the recurrent neural network to better capture the momentum effect in stock market returns. It is beyond the focus of our paper to fully examine the sources of portfolio gains from deep learning methods. Exploring this question in detail, following the approach of \cite{nagel2025seemingly}, presents an interesting avenue for future research.

Panel B in Table \ref{tab: portfolio returns cer and sr} shows the performance metrics across expansions and recessions for annual rebalancing. The nonlinear methods continue to generate stronger portfolio performance relative to linear models. Also, deep learning models tend to generate lower CERs and SRs for annual rebalancing, especially in recessions. This demonstrates that the benefits of nonlinear models remain substantial even with less frequent rebalancing, although these benefits are less pronounced. This is because the portfolio manager would naturally benefit from adjusting positions more often during turbulent periods. Interestingly, at the annual rebalancing horizon, the LSTM tends to underperform NN1–NN3, particularly during adverse periods. This contradicts the ranking reported in Panel A. This pattern suggests that in our low-dimensional market-timing setting, the additional structure provided by the recurrent architecture is most useful for exploiting short- and medium-term dynamics that are important when allocations are revised quarterly. However, when decisions are updated only annually, much of this incremental information is averaged out. This interpretation is consistent with evidence that LSTMs can capture long-term dependencies in richer SDF-based environments \citep{chen2020deep}, but suggests that very long lags are not the main drivers of economic value in the specific portfolio problem under consideration.

\begin{table}[t!]
\centering
\footnotesize
\caption{Portfolio Return Statistics}
\begin{minipage}{\textwidth} 
This table reports mean, standard deviation, skewness, and kurtosis of optimal portfolio returns for different models as outlined in Section \ref{section: optimal portfolios}. All statistics are expressed in monthly terms. Panel A shows the results for the case when the investor maximizes a 6-month portfolio return and changes the allocations quarterly. Panel B shows the results for a 2-year horizon and annual rebalancing. The portfolio construction starts in February 1955. The sample period spans from January 1945 to December 2018.
\end{minipage}
\vspace{10pt}

\begin{tabular}{lccccccccc}
\toprule
& \multicolumn{1}{c}{EH} & \multicolumn{1}{c}{OLS1} & \multicolumn{1}{c}{OLS2} & \multicolumn{1}{c}{OLS3} & \multicolumn{1}{c}{OLS4} & \multicolumn{1}{c}{NN1} & \multicolumn{1}{c}{NN2} & \multicolumn{1}{c}{NN3} & \multicolumn{1}{c}{LSTM} \\
\midrule
\multicolumn{10}{l}{Panel A: 6-month horizon  and quarterly rebalancing}      \\
\midrule
Mean  & 0.937 & 2.213 & 4.138 & 4.676 & 6.762 & 10.728 & 9.605 & 9.533 & 11.715 \\
St.dev. & 5.504 & 13.871 & 19.122 & 15.353 & 20.502 & 18.601 & 17.641 & 19.121 & 19.343 \\
Skew  & -0.472 & -0.615 & -0.893 & -0.332 & -0.881 & -0.844 & -0.816 & -0.786 & -0.046 \\
Kurt  & 4.353 & 8.609 & 10.400  & 7.631 & 9.182 & 11.707 & 12.237 & 11.172 & 4.860 \\

\midrule
\multicolumn{10}{l}{Panel B: 2-year horizon and annual rebalancing}      \\
\midrule
Mean  & 0.978 & 2.184 & 3.058 & 5.093 & 4.908 & 7.333 & 5.634 & 4.445 & 7.722 \\
St.dev. & 5.849 & 14.443 & 19.189 & 17.655 & 17.512 & 15.331 & 11.937 & 9.916 & 18.87 \\
Skew  & -0.452 & -0.492 & -1.058 & -0.989 & -0.787 & -0.275 & 0.469 & 0.799 & -0.013 \\
Kurt  & 4.386 & 7.104 & 10.566 & 13.550 & 10.737 & 9.113 & 10.416 & 14.775 & 6.433 \\

\bottomrule
\end{tabular}%
\label{tab: portfolio return statistics}%
\end{table}%

Table \ref{tab: portfolio return statistics} presents additional statistics of portfolio returns for different methodologies. The models using NNs generate out-of-sample returns with significantly larger means. Intuitively, this occurs because machine learning methods excel at predicting risk premiums, that is, the conditional expectations of returns. The linear regressions and vanilla NNs do not take the time-varying volatility of returns into account. Hence, these models predict negative skewness and excess kurtosis (since they ignore a fat-tailed return distribution). Interestingly, although an LSTM network does not consider time variation in return volatility, it can identify the periods of high return variance using the long-term memory of its cells (including realized return variance as one of the predictors also helps). This results in better skewness and lower excess kurtosis. The statistics for the longer horizon portfolio are improved for the standard neural networks, where properties remain largely the same or slightly deteriorate for other models.

\begin{figure}[t!]
\footnotesize
\caption{\bf Cumulative Returns}

\centering
\vspace{10pt}
\begin{justify}
\footnotesize This figure illustrates the cumulative log returns of optimal portfolio strategies from different models outlined in Section \ref{section: optimal portfolios}. The left panel shows the results when the investor maximizes a 6-month portfolio return and changes the allocations quarterly. The right panel shows the results for a 2-year horizon and annual rebalancing. The shaded areas denote recession periods as defined by NBER. The portfolio construction starts in February 1955. The sample period spans from January 1945 to December 2018.
\end{justify}
\subfigure[6-month horizon and quarterly rebalancing]{\includegraphics[width=.5\textwidth]{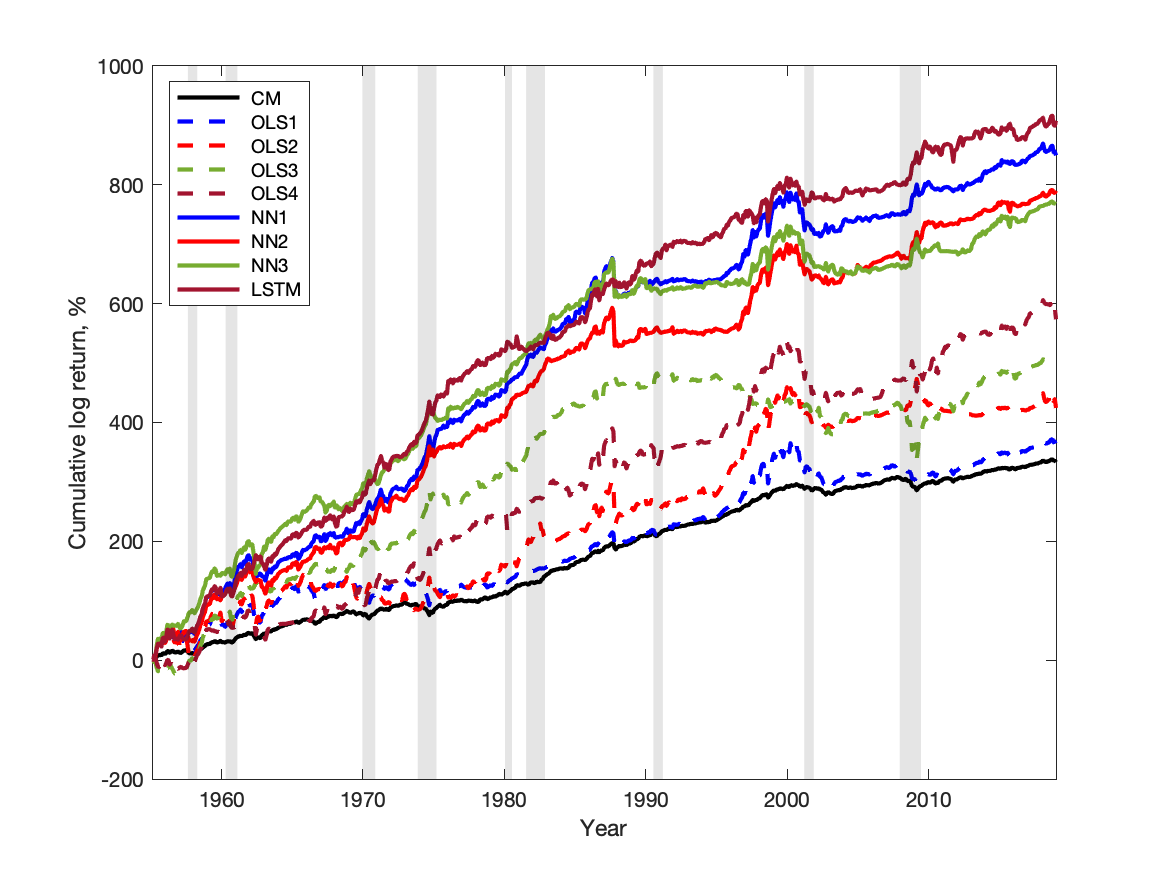}}\vspace{-1em}
\subfigure[2-year horizon and annual rebalancing]{\includegraphics[width=.5\textwidth]{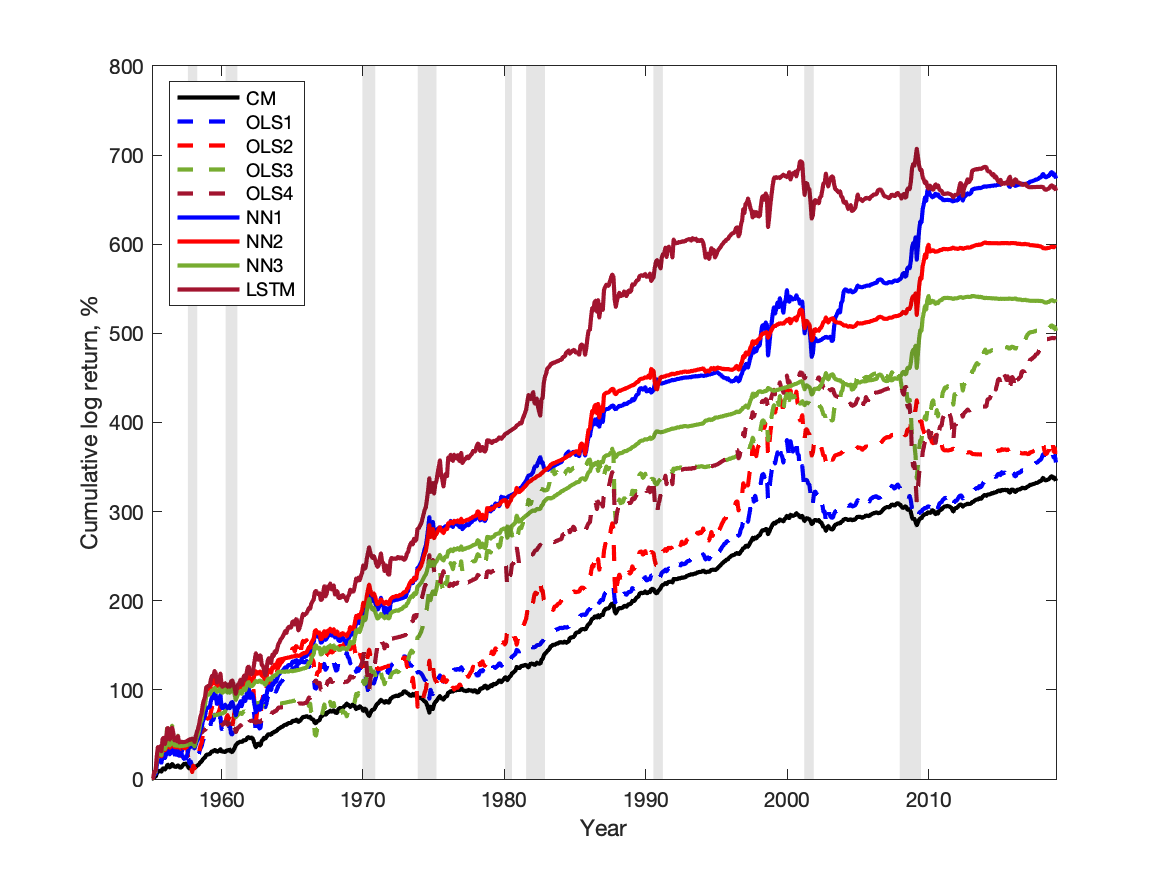}}\hspace*{-1em}
\label{fig: cumulative returns}
\end{figure}

We visually summarize the previous results in Figure \ref{fig: cumulative returns}, which shows the cumulative sum of log portfolio returns. The left panel shows that NNs outperform other models by a large margin. The LSTM dominates the remaining networks by the end of the period considered, with a particularly pronounced difference in the second half of the sample. In relation to specific historical events, all NNs produce steady positive portfolio performance during the 2007-2008 Financial Crisis. Interestingly, the LSTM network additionally avoids a largely unexpected stock market crash, Black Monday, on October 19, 1987. Figure \ref{fig: cumulative returns} also shows that weaker statistical performances for the passive strategy with annual rebalancing lead to lower cumulative returns across all models.

\section{Further Analysis}
\label{section: Further Analysis}

This section evaluates the impact of predictors on the performance of portfolios based on different models, with a particular focus on deep machine learning methods. It then analyses the performance of the portfolio returns constructed in the previous section over seven decades in the post-war period under consideration. Furthermore, this section relates the economic gains of the best-performing model to common drivers of asset prices. It shows the robustness of our conclusions to alternative measures of portfolio performance, transaction costs, borrowing, and short selling restrictions, a larger rolling window used to train machine learning methods. Finally, the last subsection implements a timing strategy.

\subsection{Variable Importance}

We now assess the contribution of various predictors to the performance of portfolios based on the NN3 and LSTM models to better understand the contribution of the time-series dimension in the deep machine learning methods. We do this by ranking each variable according to the reduction in the Sharpe ratio of an optional portfolio, which omits a particular variable as a predictor, compared to the Sharpe ratio of a benchmark portfolio, which uses all variables. Specifically, we compute the ratio $100 \times (1-SR_{j}/SR_{all})$ where $SR_{j}$ denotes the performance of a portfolio using all predictors except the $j$th variable and $SR_{all}$ denotes the performance of a portfolio using all predictors. This value measures the percentage reduction in the risk-adjusted performance when a $j$th variable is dropped. In other words, such a ratio tells us how much a variable $j$ contributes to the portfolio value and how important it is relative to others.

\begin{figure}[t!]
\caption{Variable importance}
\vspace{10pt}
\begin{justify}
\footnotesize This figure demonstrates the importance of variables for the portfolios based on the NN3 neural network and the LSTM model. The variable importance metric is calculated as the reduction in the Sharpe ratio of an optimal portfolio, which omits a particular variable as a predictor, compared to the Sharpe ratio of a benchmark portfolio, which uses all variables. The variables include twelve predictors: dividend yield (dy), log earning price ratio (ep), dividend payout ratio (de), book-to-market ratio (bm), net equity expansion (ntis), treasury bill rates (tbl), term spread (tms), default yield spread (dfy), default return spread (dfr), cross-sectional premium (csp), inflation growth (infl), and stock variance (svar).
\end{justify}
\vspace{10pt}
\centering\includegraphics[width=0.8\textwidth]{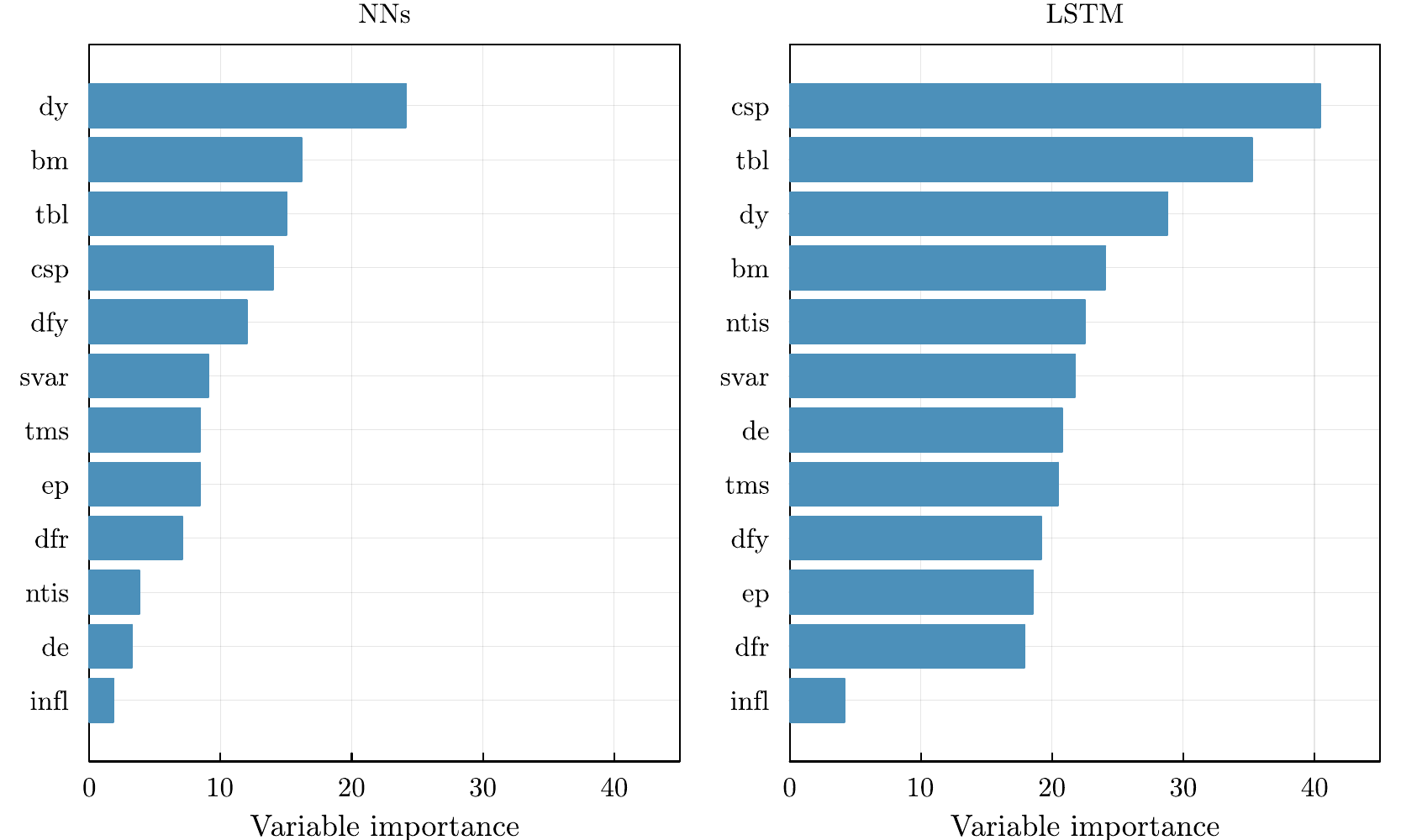}
\label{fig: varimportance}
\end{figure}

Figure \ref{fig: varimportance} demonstrates the variable importance results. Focusing on the NN3 model, the dividend yield is the strongest driver of portfolio performance, accounting for a quarter of the optimal portfolio's Sharpe ratio. This observation is not surprising because the dividend yield appears to be the strongest predictor of long-term stock market returns, and hence, this excellent long-term predictive power directly translates into significant portfolio gains. Other known long-term predictors, including the book-to-market ratio, treasury bill rates, and cross-sectional premium, contribute around 15\% of the observed Sharpe ratio. 

Turning to the LSTM model, the contribution of all variables except inflation is amplified. Intuitively, these variables exhibit time series dynamics relevant to the forecasting task that the LSTM can explore. Most prominently, the three predictors -- cross-sectional premium, treasury bill rates, and the dividend yield -- contribute around 40\%, 35\%, and 30\% to the observed performance. Notably, the contribution of others is high, around 20\%, which is comparable to the dividend yield in the NN3 model. Interestingly, four of the five most significant predictors (dy, tbl, and csp) in the LSTM model also belong to the set of five predictors with potentially appealing characteristics identified by \cite{welch2007comprehensive}.\footnote{In fact, Section 5.4 in \cite{welch2007comprehensive} focuses on a set of seven most appealing predictors, which additionally includes percent equity issuing (eqis) and wealth-to-consumption ratio (cay3). However, these two additional variables are unavailable at the monthly frequency.}

In sum, the two neural networks generally agree on the predictors' relative importance. These results align well with the previous findings on the most important predictors documented in \cite{welch2007comprehensive}. Those variables that are relevant in linear regressions continue to be the most important in nonlinear models. Furthermore, although NN3 has already introduced the nonlinear features, all variables yield higher contributions to the Shape ratios in LSTM. This indicates that the time series dependencies captured in LSTM provide incremental value to the nonlinear properties present in NN3.

\begin{figure}[t!]
\caption{Nonlinear dependence in variables}
\vspace{10pt}
\begin{justify}
\footnotesize This figure reports ranks (1 = highest dependence) of the twelve monthly Welch--Goyal predictors — dividend yield (dy), log earnings price ratio (ep), dividend payout ratio (de), book-to-market ratio (bm), net equity expansion (ntis), stock variance (svar), treasury bill rate (tbl), cross-sectional premium (csp), term spread (tms), default yield spread (dfy), default return spread (dfr), and inflation (infl). Ranks are computed using three serial-dependence measures over lags $k=1,\dots,24$: (i) the average autocorrelation (left); (ii) the \emph{distance correlation} between $X_t$ and $X_{t-k}$ averaged across lags (middle), which captures general (possibly nonlinear) dependence and equals zero under independence; and (iii) a \emph{Bartlett-weighted} average distance correlation (right), which down-weights longer lags with weights $w_k = 1 - k/25$. Higher values indicate stronger short- and medium-run serial structure; the weighted version emphasizes short-horizon dynamics, while the unweighted average reflects overall persistence across the first 24 lags.
\end{justify}
\vspace{10pt}
\centering\includegraphics[width=\textwidth]{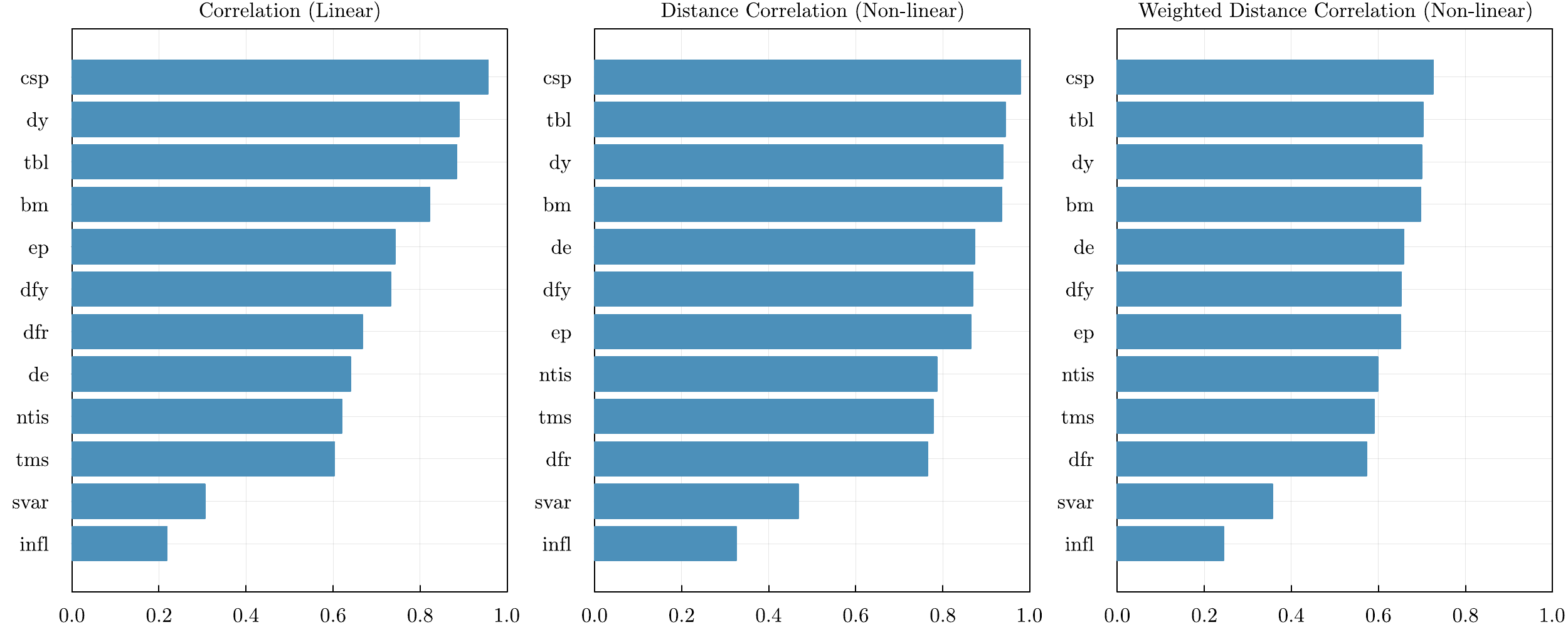}
\label{fig: correlations}
\end{figure}

To further understand the importance of the time series dependence explored in LSTM, Figure \ref{fig: correlations} ranks the variables as follows: (i) the average of the first 24 autocorrelations (left panel); (ii) the average distance correlation between the 24 lags (middle panel); and (iii) a short-horizon version which applies Bartlett weights to reduce the influence of longer lags (right panel). Distance correlation is a non-parametric dependence measure that captures nonlinear dynamics and is zero under independence. Aggregating it over lags summarises each series’ overall serial dependence. A small set of predictors — csp, tbl, dy and bm — rise to the top across all three panels, while infl and svar sit at the bottom. The weighted and unweighted non-parametric rankings are nearly identical, indicating that the leading predictors remain significant at both short and medium horizons.
These rankings align closely with the internal importance diagnostics of our models. Figure \ref{fig: varimportance} shows that both the LSTM and the deep feedforward networks attribute substantial importance to csp, tbl and dy, with bm and other persistent series close behind. Therefore, the variables identified by our nonlinear dependence measures as being the most serially dependent are also those driving the economic performance of all neural networks, not just the LSTM. Since LSTMs are expressly designed to learn non-linear long- and short-run temporal dependencies through gated memory, it is understandable that the recurrent architecture relies heavily on such predictors. This may explain its incremental gains in certain environments compared to feedforward networks lacking explicit sequence memory.

However, Figure \ref{fig: correlations} alone cannot conclusively prove that the advantage of LSTMs arises solely from temporal dependence: the differences in dependence measures across variables are moderate and the ranking of predictors is similar across models. Therefore, we interpret the dependence evidence as consistent with, rather than as conclusive proof of, a time-series channel behind the economic value of deep learning in our application. As these measures are computed in a moving window, the ranking and degree of persistence are both time-varying, which reinforces the idea that the networks adapt to an evolving time-series structure.

\subsection{Subsample Analysis}

We start by examining whether examine whether the superior portfolio performance implied by neural networks varies in subsamples other than expansions and recessions. Table \ref{tab: subsamples} presents the certainty-equivalent yields and Sharpe ratios computed for each decade in our sample. Regarding the CERs, we reiterate the paper's main finding: all neural networks outperform the expectations hypothesis model, with the LSTM often among the strongest performers. Specifically, except for the last decade, the table shows that the LSTM network produces CERs that exceed those implied by the no-predictability framework. A formal test indicates that the improvement of the LSTM network over the EH model is statistically significant during the first three decades. In later periods, however, higher CERs are statistically comparable to those from the EH model, and other deep networks typically deliver gains of a similar magnitude.

The linear models perform well across the 1990s and 2010s, during which the stock market grew steadily. Also, the rolling-window linear regressions tend to perform better than those using the expanding-window estimation, emphasizing the role of time-varying betas and changing information sets. For instance, \cite{welch2007comprehensive} show that dividend yield exhibited a strong predictive power for stock market returns from 1970 to mid-1990, with a weaker but mostly positive out-of-sample performance during the first two decades after World War II. In contrast, it produced large prediction errors during the 1995-2000 period and the 2000s. As a result, Table \ref{tab: subsamples} shows that the OLS3 model generates high CERs from 1955 to 1989, exhibiting statistically better performance than EH in some cases, but the model is weaker in later years when the forecast based on dividend yield had strong underperformance.

\begin{table}[t!]
\footnotesize
\centering
\caption{Portfolio Performance across Subsamples}
\begin{minipage}{\textwidth} 
This table reports the annualized certainty equivalent returns and monthly Sharpe ratios for different models outlined in Section \ref{section: optimal portfolios}. The table shows the results when the investor maximizes a 6-month portfolio return and changes the allocations quarterly. The table computes the statistics for each of the seven decades since WWII. For the statistical significance of CERs, we report a one-sided p-value (in parentheses) of the test \'{a} la \cite{diebold1995comparing}. In particular, we regress the difference in utilities for each model
$\mathcal{M}_s$
and EH:
$\mathcal{U}_{t+T}^{\mathcal{M}_s} - \mathcal{U}_{t+T}^{EH} = \alpha^{\mathcal{M}_s} + \varepsilon_{t+T},$
where 
$\mathcal{U}_{t+T}^{X} = \frac{\left(r_{p,t+T}^{X}\right)^{1-\gamma}}{1-\gamma}$
and
$r_{p,t+T}^{X}$
is the cumulative portfolio return with the horizon 
$T.$
Testing for the difference in the CERs boils down to a test for the significance in 
$\alpha^{\mathcal{M}_s}.$
We flag in \textbf{bold font} CER values that are significant at the 10\% confidence level. The portfolio construction starts in February 1955. The sample period spans from January 1945 to December 2018.
\end{minipage}
\vspace{10pt}

\begin{tabular}{lccccccccc}
\toprule
& \multicolumn{1}{c}{EH} & \multicolumn{1}{c}{OLS1} & \multicolumn{1}{c}{OLS2} & \multicolumn{1}{c}{OLS3} & \multicolumn{1}{c}{OLS4} & \multicolumn{1}{c}{NN1} & \multicolumn{1}{c}{NN2} & \multicolumn{1}{c}{NN3} & \multicolumn{1}{c}{LSTM} \\
\midrule
1955-1959 &       &       &       &       &       &       &       &       &  \\
[0.3em]

CER   & 5.467 & 4.376 & 3.219 & 9.495 & 5.545 & \textbf{15.611} & \textbf{12.220} & \textbf{23.120} & \textbf{15.455} \\
p-value &       & (0.631) & (0.707) & (0.149) & (0.490) & (0.000) & (0.017) & (0.000) & (0.001) \\
SR    & 0.225 & 0.188 & 0.209 & 0.258 & 0.154 & 0.319 & 0.278 & 0.431 & 0.341 \\
\midrule
1960-1969 &       &       &       &       &       &       &       &       &  \\
[0.3em]

CER   & 4.197 & 0.580 & -4.193 & \textbf{8.354} & 0.619 & \textbf{7.498} & \textbf{7.608} & 5.627 & \textbf{14.015} \\
p-value &       & (0.959) & (0.991) & (0.024) & (0.931) & (0.094) & (0.042) & (0.360) & (0.000) \\
SR    & 0.062 & 0.064 & 0.030 & 0.157 & 0.067 & 0.164 & 0.148 & 0.181 & 0.241 \\
\midrule
1970-1979 &       &       &       &       &       &       &       &       &  \\
[0.3em]

CER   & 3.312 & 0.599 & 0.223 & \textbf{9.275} & \textbf{8.580} & \textbf{17.750} & \textbf{15.690} & \textbf{13.618} & \textbf{20.658} \\
p-value &       & (1.000) & (0.847) & (0.015) & (0.026) & (0.000) & (0.000) & (0.000) & (0.000) \\
SR    & -0.107 & -0.097 & 0.005 & 0.149 & 0.180 & 0.274 & 0.248 & 0.224 & 0.309 \\
\midrule
1980-1989 &       &       &       &       &       &       &       &       &  \\
[0.3em]

CER   & 9.215 & 7.243 & -3.130 & 11.276 & -1.450 & 3.315 & 1.216 & 2.603 & 10.241 \\
p-value &       & (0.992) & (0.983) & (0.148) & (0.969) & (0.820) & (0.906) & (0.864) & (0.282) \\
SR    & 0.048 & -0.005 & 0.072 & 0.139 & 0.055 & 0.166 & 0.123 & 0.142 & 0.112 \\
\midrule
1990-1999 &       &       &       &       &       &       &       &       &  \\
[0.3em]

CER   & 8.101 & \textbf{11.808} & \textbf{13.704} & -4.393 & \textbf{12.085} & \textbf{10.817} & \textbf{10.301} & 6.429 & 9.815 \\
p-value &       & (0.002) & (0.002) & (1.000) & (0.039) & (0.052) & (0.099) & (0.842) & (0.202) \\
SR    & 0.222 & 0.185 & 0.219 & -0.168 & 0.218 & 0.168 & 0.172 & 0.100 & 0.150 \\
\midrule
2000-2009 &       &       &       &       &       &       &       &       &  \\
[0.3em]

CER   & 0.000 & -7.445 & -7.091 & -15.834 & -10.520 & -3.889 & -0.205 & -7.991 & 1.726 \\
p-value &       & (1.000) & (0.990) & (0.998) & (0.998) & (0.869) & (0.531) & (0.995) & (0.238) \\
SR    & -0.091 & -0.125 & -0.055 & -0.039 & -0.040 & 0.028 & 0.060 & -0.062 & 0.084 \\
\midrule
2010-2018 &       &       &       &       &       &       &       &       &  \\
[0.3em]

CER   & 4.230 & \textbf{6.268} & 0.997 & \textbf{9.757} & \textbf{9.508} & \textbf{6.642} & \textbf{5.690} & \textbf{7.241} & 2.074 \\
p-value &       & (0.000) & (0.999) & (0.000) & (0.000) & (0.004) & (0.049) & (0.023) & (0.842) \\
SR    & 0.237 & 0.220 & -0.020 & 0.255 & 0.161 & 0.150 & 0.175 & 0.213 & 0.096 \\
\bottomrule
\end{tabular}%
\label{tab: subsamples}%
\end{table}%

Turning to the SRs, NNs provide the investor with substantially higher Sharpe ratios except in the 1990s and 2010s, when they perform slightly worse. These results are consistent with our previous findings. Indeed, the U.S. stock market was strongly bullish in these two decades, which were marked by prolonged stock market expansions. In contrast, the Black Monday crash occurred in 1987 and the S\&P 500 index recovered slowly, only by the end of the 1980s. Further, the beginning of the new millennium experienced two major crashes driven by the burst of the dot-com bubble and the subprime mortgage crisis. Table \ref{tab: subsamples} shows that NNs perform significantly better than other specifications during decades with major stock bear markets and provide statistically equal results during bull markets, which is consistent with our previous results across expansions and recessions. 

\subsection{Drivers of Portfolio Performance}
This section explores the link between the economic gains implied by the best-performing machine learning framework and prominent drivers of asset prices. In particular, we focus on the portfolio choice problem of the investor with a 6-month investment horizon and quarterly rebalancing who uses the LSTM to forecast future stock market returns. Formally, we establish this link by running a set of univariate regressions of the investor's utility from future portfolio returns on the set of structural determinants of risk premia.

Our choice of variables is motivated by existing studies. For instance, a large strand of the literature (see \cite{buraschi2007habit} and \cite{dumas2009equilibrium} among others) emphasizes the importance of disagreement for asset prices. In our analysis, we employ the Survey of Professional Forecasters to proxy for real disagreement ($DiB(g)$) and nominal disagreement ($DiB(\pi)$), which are constructed as the interquartile range of 6-month-ahead forecasts of GDP and CPI growth. Motivated by the well-established link between asset prices and uncertainty, we employ a novel measure of economic uncertainty ($UNbex$) constructed from financial variables at high frequencies \citep{bekaert2019time}. 

\begin{table}[t!]
\centering
\footnotesize
\caption{Drivers of Portfolio Performance}
\begin{minipage}{\textwidth} 
This table reports the regression estimates, Newey-West p-values (in parentheses) and $R^2$ of economic gains on a set of selected variables determining risk premia. Economic gains are computed for portfolio returns for the best performing model employing the LSTM prediction of stock market returns. The independent variables proxy for real disagreement $DiB(g),$ nominal disagreement $DiB(\pi),$ economic uncertainty $UNbex,$ risk aversion via consumption growth $-Surplus$ or financial variables $RAbex,$ the VIX index $VIX,$ and realized stock market volatility $\sigma.$ The variables on the left and right sides are standardized. We flag in \textbf{bold font} regression estimates that are significant at the 10\% confidence level.
\end{minipage}
\vspace{10pt}

\begin{tabular}{ccccccccc}
\toprule
& $DiB(g)$ & $DiB(\pi)$ & $UNbex$ & $-Surplus$ & $RAbex$ & $VIX$   & $\sigma$ & $R^2 (\%)$ \\
\midrule
    (i)   & \textbf{0.198} &       &       &       &       &       &       & 3.907 \\
& (0.002) &       &       &       &       &       &       &  \\
(ii)  &       & \textbf{0.188} &       &       &       &       &       & 3.523 \\
&       & (0.019) &       &       &       &       &       &  \\
(iii) &       &       & 0.115 &       &       &       &       & 1.318 \\
&       &       & (0.127) &       &       &       &       &  \\
(iv)  &       &       &       & 0.037 &       &       &       & 0.141 \\
&       &       &       & (0.561) &       &       &       &  \\
(v)   &       &       &       &       & 0.098 &       &       & 0.962 \\
&       &       &       &       & (0.144) &       &       &  \\
(vi)  &       &       &       &       &       & 0.104 &       & 1.084 \\
&       &       &       &       &       & (0.177) &       &  \\
(vii) &       &       &       &       &       &       & 0.004 & 0.002 \\
&       &       &       &       &       &       & (0.908) &  \\
\bottomrule
\end{tabular}%
\label{tab:drivers}%
\end{table}%

We next examine the relationship between portfolio gains and the time-varying risk aversion of investors. Following \cite{wachter2006consumption}, we approximate risk aversion via the negative weighted average of consumption growth rates over a moving window of 10 years ($- Surplus$). We compare the results to an alternative measure of risk aversion extracted from financial variables \citep{bekaert2019time}. Finally, we relate portfolio utilities to stock market volatility by using the risk-neutral volatility ($VIX$) as measured by the VIX index and by using the realized volatility ($\sigma$) as measured by the root of the intra-month sum of squared daily S\&P500 returns.

Table \ref{tab:drivers} presents the regression results. The relationship between future realized portfolio gains and most structural risk factors is relatively weak. Indeed, we document that only dispersion in beliefs about real or nominal growth is positively and statistically significantly linked to the investor's utilities. Intuitively, this result is expected since machine learning methods significantly outperform competing models during recessionary periods, when uncertainty and disagreement in forecasts are large. The third panel in Table \ref{tab:drivers}  further confirms this positive association between economic uncertainty and portfolio gains, however, the link is statistically weaker compared to disagreement measures. We obtain positive coefficients on the remaining risk factors except for realized stock market volatility.

\subsection{Alternative Measures of Performance}

Although certainty equivalent yields and Sharpe ratios are common measures of portfolio performance considered in the literature, the investor may use alternative statistics to evaluate their investment strategies, including maximum drawdown, maximum one-month loss, and average monthly turnover. For each model $\mathcal{M}_s$, we define maximum drawdown
\begin{equation}
\text{Max DD} = \max\limits_{t_0\leq t_1\leq t_2 \leq T_0} \Big[\hat{r}_{t_0}^{t_1,\mathcal{M}_s}-\hat{r}_{t_0}^{t_2,\mathcal{M}_s}\Big],
\end{equation}
in which 
$\hat{r}_{t_0}^{t,\mathcal{M}_s}$
denotes the cumulative portfolio return from time
$t_0$
through
$t,$
while 
$t_0$
and
$T_0$
are the months of the first and last predictions. The maximum one-month loss measures the largest portfolio decline during the period considered. The average monthly turnover is defined as
\begin{equation}
\text{Turnover} = \frac{1}{T_0 - t_0}\sum_{t=t_0+1}^{T_0}\Big| \omega_{t} - \omega_{t-1}\cdot \hat{r}_{t-1}^{\mathcal{M}_s} \Big|,
\end{equation}
where
$\omega_{t-1}$
is the weight of the stock index.

Table \ref{tab: alternative measure of performance} shows the results for alternative performance statistics. We first focus on actively managed portfolios with quarterly rebalancing and then move to more passive investment strategies with annual rebalancing. The maximum drawdown experienced by NN1 through NN3 is between 68\% and 83\% on a monthly basis. The linear models predict comparable or even larger drawdowns, whereas the constant mean and constant volatility model delivers a mild loss of around 23\%. In contrast, the maximum drawdown for LSTM is around 46\%, the mildest decline among the predictive models. Panel A further shows a similar picture for the maximum one-month loss of the portfolio: linear models and NNs tend to generate the worst one-period performance, while the LSTM strategy experiences a milder loss. Thus, the LSTM specification is the most successful in avoiding large losses over short- and long-term periods, even though it comes at the expense of the higher turnover. 

\begin{table}[t!]
\centering
\footnotesize
\caption{Drawdowns, Maximum Loss, and Turnover}
\begin{minipage}{\textwidth} 
This table reports alternative out-of-sample performance measures --- maximum drawdown, maximum 1-month loss, and turnover --- of optimal portfolio returns for different methodologies used to predict future S\&P 500 excess returns as outlined in Section \ref{section: optimal portfolios}. All statistics are expressed in percentages. Panel A shows the results when the investor maximizes a 6-month portfolio return and changes the allocations quarterly. Panel B demonstrates the results for a 2-year horizon and annual rebalancing. The portfolio construction starts in February 1955. The sample period spans from January 1945 to December 2018.
\end{minipage}
\vspace{10pt}

\begin{tabular}{lccccccccc}
\toprule
& \multicolumn{1}{c}{EH} & \multicolumn{1}{c}{OLS1} & \multicolumn{1}{c}{OLS2} & \multicolumn{1}{c}{OLS3} & \multicolumn{1}{c}{OLS4} & \multicolumn{1}{c}{NN1} & \multicolumn{1}{c}{NN2} & \multicolumn{1}{c}{NN3} & \multicolumn{1}{c}{LSTM} \\
\midrule
\multicolumn{10}{l}{Panel A: 6-month horizon and quarterly rebalancing}      \\
\midrule
Max DD & 22.795 & 76.236 & 74.251 & 144.760 & 100.956 & 74.572 & 68.248 & 82.72 & 45.995 \\
Max 1M Loss & 7.795 & 33.325 & 57.974 & 31.756 & 57.974 & 57.974 & 57.974 & 57.974 & 35.011 \\
Turnover & 0.506 & 4.286 & 10.968 & 17.890 & 34.531 & 23.008 & 23.407 & 29.616 & 32.814 \\
\midrule
\multicolumn{10}{l}{Panel B: 2-year horizon and annual rebalancing}      \\
\midrule
Max DD & 25.002 & 90.896 & 83.370 & 123.279 & 145.663 & 74.972 & 34.562 & 25.136 & 64.433 \\
Max 1M Loss & 8.036 & 29.431 & 57.974 & 57.974 & 38.862 & 34.375 & 24.419 & 25.136 & 35.011 \\
Turnover & 0.584 & 4.289 & 8.804 & 10.773 & 11.329 & 11.352 & 8.725 & 6.771 & 17.459 \\
\bottomrule
\end{tabular}%
\label{tab: alternative measure of performance}%
\end{table}%

Panel B in Table \ref{tab: alternative measure of performance} shows that the investor engaging in less frequent portfolio rebalancing is generally less efficient in forming the optimal portfolio if he relies on the linear regressions. Interestingly, the benefits of deep learning methods remain similar and even improve in some cases. For instance, the maximum one-month and drawdown losses tend to increase from 83\% to more than 140\% for the linear models, while NNs produce the largest declines, from 25\% to 35\% per month. Furthermore, as the portfolio weights are kept unchanged for longer investment periods, the turnover is reduced. Thus, the passive investor who is mainly interested in reducing his short- and long-term tail risks would still find NNs useful, while he does not benefit from linear predictive models.

In sum, exploiting expected return predictability via NNs for portfolio construction leads to riskier investments. It also generates increased turnover, especially for the best-performing model using the LSTM network. A natural question arises whether transaction costs offset these benefits.

\subsection{Portfolio Performance with Transaction Costs}
This subsection extends the main analysis by accounting for the effect of transaction costs. Specifically, we consider low and high transaction costs that are equal to the percentage paid by the investor for the change in value traded. Let $\tau$ denote a transaction cost parameter. Then the transaction-cost-adjusted returns are defined as
$$
\hat{r}_{t}^{\tau,\mathcal{M}_s} = \hat{r}_{t}^{\mathcal{M}_s} - \tau \big|\omega_{t} - \omega_{t-1}\cdot \hat{r}_{t-1}^{\mathcal{M}_s} \big|,
$$
where
$\tau$
can attain one of the two possible values
$\tau_l = 0.1\%$
or 
$\tau_h = 0.5\%.$

\begin{table}[t!]
\centering
\footnotesize
\caption{Portfolio Performance with Transaction Costs}
\begin{minipage}{\textwidth} 
This table reports the annualized certainty equivalent returns and Sharpe ratios for different models outlined in Section \ref{section: optimal portfolios}. The top and bottom sections of the table compute optimal returns with low $(\tau = 0.1\%)$ and high $(\tau = 0.5\%)$ transaction costs. Panels A and C show the results when the investor maximizes a 6-month portfolio return and changes the allocations quarterly. Panel B and D show the results for a 2-year horizon and annual rebalancing. Each panel computes the statistics for the whole sample. For the statistical significance of CERs, we report a one-sided p-value (in parentheses) of the test \'{a} la \cite{diebold1995comparing}. In particular, we regress the difference in utilities for each model
$\mathcal{M}_s$
and EH:$\mathcal{U}_{t+T}^{\mathcal{M}_s} - \mathcal{U}_{t+T}^{EH} = \alpha^{\mathcal{M}_s} + \varepsilon_{t+T},$
where 
$\mathcal{U}_{t+T}^{X} = \frac{\left(r_{p,t+T}^{X}\right)^{1-\gamma}}{1-\gamma}$
and
$r_{p,t+T}^{X}$
is the cumulative portfolio return with the horizon 
$T.$
Testing for the difference in the CERs boils down to a test for the significance in 
$\alpha^{\mathcal{M}_s}.$
We flag in \textbf{bold font} those CER values that are significant at the 10\% confidence level. The portfolio construction starts in February 1955. The sample period spans from January 1945 to December 2018.
\end{minipage}
\vspace{10pt}

\begin{tabular}{lccccccccc}
\toprule
& \multicolumn{1}{c}{EH} & \multicolumn{1}{c}{OLS1} & \multicolumn{1}{c}{OLS2} & \multicolumn{1}{c}{OLS3} & \multicolumn{1}{c}{OLS4} & \multicolumn{1}{c}{NN1} & \multicolumn{1}{c}{NN2} & \multicolumn{1}{c}{NN3} & \multicolumn{1}{c}{LSTM} \\
\midrule
\multicolumn{10}{c}{Low Transaction Costs}\\
\midrule
\multicolumn{10}{l}{Panel A: 6-month horizon and quarterly rebalancing}      \\
[0.5em]

CER    & 4.731 & 2.589 & -0.171 & 2.548 & 2.036 & \textbf{6.996} & \textbf{6.689} & 5.115 & \textbf{9.592} \\
p-value &       & (1.000) & (1.000) & (0.953) & (0.978) & (0.045) & (0.054) & (0.391) & (0.000) \\
[0.5em]

SR    & 0.049 & 0.045 & 0.060 & 0.084 & 0.089 & 0.162 & 0.153 & 0.139 & 0.169 \\
\midrule
\multicolumn{10}{l}{Panel B: 2-year horizon and annual rebalancing}      \\
[0.5em]

CER    & 4.535 & 0.998 & -0.085 & 0.765 & -0.241 & \textbf{6.191} & \textbf{6.783} & 6.366 & \textbf{5.413} \\
p-value &       & (1.000) & (1.000) & (0.999) & (0.998) & (0.001) & (0.000) & (0.000) & (0.033) \\
[0.5em]

SR    & 0.048 & 0.043 & 0.044 & 0.081 & 0.079 & 0.135 & 0.134 & 0.127 & 0.115 \\
\midrule
\multicolumn{10}{c}{High Transaction Costs}\\
\midrule
\multicolumn{10}{l}{Panel C: 6-month horizon and quarterly rebalancing}      \\
[0.5em]

CE    & 4.706 & 2.370 & -0.736 & 1.609 & 0.193 & 5.791 & 5.501 & 3.592 & \textbf{7.910} \\
p-value &       & (1.000) & (1.000) & (0.990) & (0.999) & (0.214) & (0.263) & (0.784) & (0.000) \\
[0.5em]

SR    & 0.048 & 0.041 & 0.053 & 0.068 & 0.066 & 0.145 & 0.134 & 0.117 & 0.145 \\
\midrule
\multicolumn{10}{l}{Panel D: 2-year horizon and annual rebalancing}      \\
[0.5em]

CE    & 4.506 & 0.717 & -0.586 & 0.129 & -0.943 & \textbf{5.579} & \textbf{6.396} & \textbf{6.080} & 4.563 \\
p-value &       & (1.000) & (1.000) & (1.000) & (0.999) & (0.022) & (0.000) & (0.000) & (0.453) \\
[0.5em]

SR    & 0.047 & 0.039 & 0.038 & 0.073 & 0.070 & 0.125 & 0.123 & 0.117 & 0.102 \\
\bottomrule
\end{tabular}%
\label{tab: transaction costs}%
\end{table}%

Table \ref{tab: transaction costs} summarises the out-of-sample portfolio returns for low and high transaction costs. The results show that: (i) portfolio performance decreases monotonically with increasing transaction costs; and (ii) the main conclusion of the paper remains robust: all NN specifications deliver substantially higher CERs and Sharpe ratios than traditional linear regressions and the expectations hypothesis. For quarterly rebalancing (Panels A and C), NN1 and LSTM typically generate the highest CERs and SRs. In contrast, for annual rebalancing (Panels B and D), the three deep networks perform similarly, with no single architecture dominating consistently. In some cases, NN2 or NN3 slightly outperform the recurrent specification, highlighting that our key message is about the value of deep neural networks as a whole, rather than the superiority of any individual architecture. Quantitatively, the annualized CERs for all NNs decline by less than 0.5\% and 2.1\% for the low and high transaction cost parameters, respectively. In terms of SRs, the decline in performance never exceeds 2\% and 3\% on a monthly basis for basic NNs and LSTM. However, despite a slightly detrimental effect of transaction costs, the best-performing models (NN1 and LSTM) with an actively managed portfolio generate more than two- and three-fold increases in the CERs and SRs compared to the scenario in which expected return predictability is ignored. The formal test shows that the CER gains are also statistically significant.

\subsection{Borrowing and Short-selling Constraints}

We consider an additional robustness check of the alternative assumptions about the portfolio weights. The main analysis allows the investor to borrow the money or to short-sell the stock by considering the weights in the interval 
$-1\leq\omega_{t}\leq2.$ 
In this subsection, we perform a two-step analysis: we first impose borrowing constraints by restricting the optimal weight on the risk-free investment to be non-negative, and then additionally impose short-selling constraints on the weights 
$0\leq\omega_{t}\leq1.$ 

Table \ref{tab: borrowing and short-selling constraint} reports the results for the two scenarios. We focus on the quarterly rebalancing case reported in Panels A and C. The corresponding results for the passive portfolios, which are shown in Panels B and D, remain qualitatively similar. Several observations are noteworthy. First, winsorizing the weights to narrower intervals leads to ambiguous conclusions about the performance of linear predictive models. On the one hand, the constraints prevent optimal investments, leading to smaller out-of-sample Sharpe ratios. On the other hand, using the certainty equivalent as a measure of portfolio performance, the linear specifications consistently generate improved results, with the CERs above 3.5\% in all cases. Thus, constraints on the optimal weights result in higher CERs. The reason for this seemingly counterintuitive result is that such restrictions prevent the expected utility from achieving unbounded large values \citep{johannes2014sequential} and, therefore, avoid extreme investments based on unstable predictions of linear regressions \citep{welch2007comprehensive}. Since the certainty equivalent measure takes tail behaviour of returns into account, less extreme investments ultimately yield improved results.

\begin{table}[t!]
\centering
\footnotesize
\caption{Portfolio Performance with Borrowing and Short-Selling Constraints}
\begin{minipage}{\textwidth} 
This table reports the annualized certainty equivalent returns and Sharpe ratios for different models outlined in Section \ref{section: optimal portfolios}. The top section of the table imposes borrowing constrains, while the bottom section additionally assumes short-selling constraints. Panels A and C show the results when the investor maximizes a 6-month portfolio return and changes the allocations quarterly. Panels B and D show the results for a 2-year horizon and annual rebalancing. For the statistical significance of CERs, we report a one-sided p-value (in parentheses) of the test \'{a} la \cite{diebold1995comparing}. In particular, we regress the difference in utilities for each model
$\mathcal{M}_s$
and EH: $\mathcal{U}_{t+T}^{\mathcal{M}_s} - \mathcal{U}_{t+T}^{EH} = \alpha^{\mathcal{M}_s} + \varepsilon_{t+T},$
where 
$\mathcal{U}_{t+T}^{X} = \frac{\left(r_{p,t+T}^{X}\right)^{1-\gamma}}{1-\gamma}$
and
$r_{p,t+T}^{X}$
is the cumulative portfolio return with the horizon 
$T.$
Testing for the difference in the CERs boils down to a test for the significance in 
$\alpha^{\mathcal{M}_s}.$
We flag in \textbf{bold font} CER values that are significant at the 10\% confidence level. The portfolio construction starts in February 1955. The sample period spans from January 1945 to December 2018.
\end{minipage}
\vspace{10pt}

\begin{tabular}{lccccccccc}
\toprule
& \multicolumn{1}{c}{EH} & \multicolumn{1}{c}{OLS1} & \multicolumn{1}{c}{OLS2} & \multicolumn{1}{c}{OLS3} & \multicolumn{1}{c}{OLS4} & \multicolumn{1}{c}{NN1} & \multicolumn{1}{c}{NN2} & \multicolumn{1}{c}{NN3} & \multicolumn{1}{c}{LSTM} \\
\midrule
\multicolumn{10}{c}{Borrowing Constraint}\\
\midrule
\multicolumn{10}{l}{Panel A: 6-month horizon and quarterly rebalancing}      \\
[0.5em]

CER   & 4.737 & 3.662 & 3.371 & 4.149 & 4.560 & \textbf{9.122} & \textbf{7.632} & \textbf{6.900} & \textbf{8.560} \\
p-value &       & (0.999) & (0.958) & (0.770) & (0.591) & (0.000) & (0.000) & (0.003) & (0.000) \\
[0.5em]

SR    & 0.049 & 0.046 & 0.061 & 0.074 & 0.080 & 0.176 & 0.146 & 0.135 & 0.157 \\

\midrule
\multicolumn{10}{l}{Panel B: 2-year horizon and annual rebalancing}      \\
[0.5em]

CER   & 4.542 & 2.780 & 2.936 & 2.974 & 1.560 & 4.964 & \textbf{5.336} & \textbf{5.275} & \textbf{5.321} \\
p-value &       & (1.000) & (1.000) & (1.000) & (1.000) & (0.147) & (0.007) & (0.008) & (0.033) \\
[0.5em]

SR    & 0.048 & 0.044 & 0.051 & 0.049 & 0.013 & 0.101 & 0.094 & 0.087 & 0.100 \\

\midrule
\multicolumn{10}{c}{Borrowing and Short-Selling Constraints}\\
\midrule

\multicolumn{10}{l}{Panel C: 6-month horizon and quarterly rebalancing}      \\
[0.5em]

CER   & 4.737 & 3.704 & 4.707 & 5.353 & 5.708 & \textbf{7.500} & \textbf{6.758} & \textbf{7.128} & \textbf{7.775} \\
p-value &       & (0.998) & (0.528) & (0.080) & (0.004) & (0.000) & (0.000) & (0.000) & (0.000) \\
[0.5em]

SR    & 0.049 & 0.047 & 0.066 & 0.093 & 0.093 & 0.146 & 0.129 & 0.138 & 0.150 \\

\midrule
\multicolumn{10}{l}{Panel D: 2-year horizon and annual rebalancing}      \\
[0.5em]

CER   & 4.542 & 2.780 & 3.757 & 4.261 & \textbf{5.010} & \textbf{5.809} & \textbf{5.320} & \textbf{5.408} & \textbf{6.117} \\
p-value &       & (1.000) & (1.000) & (0.859) & (0.004) & (0.000) & (0.000) & (0.000) & (0.000) \\
[0.5em]

SR    & 0.048 & 0.044 & 0.054 & 0.054 & 0.071 & 0.107 & 0.109 & 0.098 & 0.107 \\
\bottomrule
\end{tabular}%
\label{tab: borrowing and short-selling constraint}%
\end{table}%

Second, unlike the linear regressions, we document a negative impact of imposing borrowing and short-selling constraints on portfolio performance implied by NNs. For instance, Panels A and B  in Table \ref{tab: borrowing and short-selling constraint} demonstrate a decline in both CERs and SRs for all NNs, with a larger drop in performance measures in response to more stringent assumptions about the weights. Nevertheless, despite the weaker performance of machine learning methods, the table confirms the key results of the main analysis. Specifically, traditional predictive models barely generate a positive value for the investor, whereas there is robust statistical evidence of substantial improvements from using NNs. 

\subsection{Different Rolling Window Sizes}

The subperiod analysis presented in Table \ref{tab: subsamples} reveals a slightly declining performance of NNs by the end of the sample. In particular, the LSTM generates higher CERs than the EH model, however, the difference proves to be statistically indistinguishable over the last four decades. This raises the question of whether the evidence in this paper holds for more recent data. This subsection demonstrates that the main conclusions of this paper indeed remain intact.

Table \ref{tab: 20 year rolling window} reports summary statistics of the out-of-sample portfolio returns, which are obtained for the subperiod from February 1969 to December 2018 as in \cite{rossi2018predicting}. In relation to the models using the rolling-window estimation, we assume a 20-year horizon to assess the impact of longer history on the performance of different methodologies, particularly machine learning methods that are assumed to work better with larger samples. Notice that the quantitative predictions of this exercise are not directly comparable to the previous results due to differences in the historical data. In particular, the period from February 1969 to December 2018 is characterized by slightly weaker market performance, which ultimately translates into a less favorable opportunity set for the investor. The return statistics in Table \ref{tab: 20 year rolling window} are consistent with this intuition. The average Sharpe ratio implied by the model with no predictability shrinks to half the size of that in the benchmark analysis. The linear models experience comparable deterioration in results. 

\begin{table}[t!]
\centering
\footnotesize
\caption{Portfolio Performance from Feb 1969:02 to Dec 2018: 20-year rolling window}
\begin{minipage}{\textwidth}
This table reports the annualized certainty equivalent returns and Sharpe ratios for different models outlined in Section \ref{section: optimal portfolios}. The rolling window estimation uses 20 years of recent data. Panel A shows the results when the investor maximizes a 6-month portfolio return and changes the allocations quarterly. Panel B shows the results for a 2-year horizon and annual rebalancing. Each panel computes the statistics for the whole sample, expansion and recession periods as defined by the NBER. For the statistical significance of CERs, we report a one-sided p-value (in parentheses) of the test \'{a} la \cite{diebold1995comparing}. In particular, we regress the difference in utilities for each model
$\mathcal{M}_s$
and EH:
$\mathcal{U}_{t+T}^{\mathcal{M}_s} - \mathcal{U}_{t+T}^{EH} = \alpha^{\mathcal{M}_s} + \varepsilon_{t+T},$
where 
$\mathcal{U}_{t+T}^{X} = \frac{\left(r_{p,t+T}^{X}\right)^{1-\gamma}}{1-\gamma}$
and
$r_{p,t+T}^{X}$
is the cumulative portfolio return with the horizon 
$T.$
Testing for the difference in the CERs boils down to a test for the significance in 
$\alpha^{\mathcal{M}_s}.$
We flag in \textbf{bold font} CER values that are significant at the 10\% confidence level. The portfolio construction starts in February 1969. 
\end{minipage}
\vspace{10pt}

\begin{tabular}{lccccccccc}
\toprule
& \multicolumn{1}{c}{EH} & \multicolumn{1}{c}{OLS1} & \multicolumn{1}{c}{OLS2} & \multicolumn{1}{c}{OLS3} & \multicolumn{1}{c}{OLS4} & \multicolumn{1}{c}{NN1} & \multicolumn{1}{c}{NN2} & \multicolumn{1}{c}{NN3} & \multicolumn{1}{c}{LSTM} \\
\midrule
\multicolumn{10}{l}{Panel A: 6-month horizon and quarterly rebalancing}      \\
\midrule
1969-2018 &       &       &       &       &       &       &       &       &  \\
[0.5em]

CER   & 4.600 & 1.763 & 0.791 & 1.025 & 3.707 & \textbf{6.762} & \textbf{6.601} & \textbf{6.236} & \textbf{7.253} \\
p-value &       & 1.000 & 1.000 & 0.984 & 0.811 & 0.018 & 0.053 & 0.061 & 0.016 \\
SR    & 0.025 & 0.010 & 0.018 & 0.059 & 0.057 & 0.135 & 0.140 & 0.132 & 0.165 \\
\midrule
Expansions &       &       &       &       &       &       &       &       &  \\
[0.5em]

CER   & 5.038 & 3.158 & 2.479 & 4.551 & 5.846 & \textbf{7.237} & \textbf{7.314} & 5.673 & \textbf{6.734} \\
p-value &       & 0.999 & 0.997 & 0.694 & 0.158 & 0.008 & 0.022 & 0.335 & 0.039 \\
SR    & 0.090 & 0.059 & 0.045 & 0.070 & 0.068 & 0.139 & 0.138 & 0.141 & 0.161 \\
\midrule
Recessions &       &       &       &       &       &       &       &       &  \\
[0.5em]

CER   & 1.846 & -6.771 & -9.496 & -18.688 & -8.560 & 3.819 & 2.347 & 4.423 & \textbf{17.657} \\
p-value &       & 1.000 & 0.999 & 0.985 & 0.980 & 0.345 & 0.465 & 0.287 & 0.002 \\
SR    & -0.251 & -0.253 & -0.123 & 0.034 & 0.023 & 0.123 & 0.156 & 0.061 & 0.226 \\
\midrule
\multicolumn{10}{l}{Panel B: 2-year horizon and annual rebalancing}      \\
\midrule
1969-2018 &       &       &       &       &       &       &       &       &  \\
[0.5em]

CER   & 4.530 & 0.508 & -2.573 & -2.558 & 2.080 & \textbf{5.788} & \textbf{5.136} & \textbf{7.038} & \textbf{6.477} \\
p-value &       & (1.000) & (1.000) & (1.000) & (1.000) & (0.026) & (0.068) & (0.000) & (0.000) \\
SR    & 0.023 & 0.008 & -0.002 & 0.008 & 0.025 & 0.126 & 0.084 & 0.117 & 0.135 \\
\midrule
Expansions &       &       &       &       &       &       &       &       &  \\
[0.5em]

CER   & 4.448 & 0.246 & -2.160 & -2.174 & 3.324 & \textbf{5.465} & 4.675 & \textbf{6.739} & \textbf{6.185} \\
p-value &       & (1.000) & (1.000) & (1.000) & (0.992) & (0.076) & (0.297) & (0.000) & (0.001) \\
SR    & 0.089 & 0.059 & 0.035 & 0.008 & 0.021 & 0.108 & 0.054 & 0.100 & 0.127 \\
\midrule
Recessions &       &       &       &       &       &       &       &       &  \\
[0.5em]

CER   & 5.089 & 2.275 & -5.061 & -4.914 & -4.333 & \textbf{8.133} & \textbf{8.538} & \textbf{9.073} & \textbf{8.409} \\
p-value &       & (0.999) & (1.000) & (1.000) & (1.000) & (0.023) & (0.004) & (0.003) & (0.009) \\
SR    & -0.248 & -0.259 & -0.194 & 0.010 & 0.045 & 0.216 & 0.211 & 0.210 & 0.181 \\
\bottomrule
\end{tabular}%

\label{tab: 20 year rolling window}%
\end{table}%

For NNs with quarterly rebalancing, we document several interesting observations. First, despite a weaker stock market performance during the period considered, monthly Sharpe ratios implied by NNs decrease marginally, with the drop approximately equal to 0.01 to 0.03 relative to the main results. Second, comparing NN1 through NN3 in terms of certainty equivalent returns, NNs yield statistically the same results. Although deeper networks generate slightly lower CERs than those predicted by shallower networks, the p-values indicate that these model-based values remain in the same equivalence class. Third, the LSTM still produces significant economic gains. Specifically, the annualized certainty equivalent yield is above 7\% and monthly Sharpe ratios remain as high as 0.165. Finally, unlike weak statistical evidence of the main results with recent data, the formal test of the results in this subsection demonstrates strong statistical evidence in favor of NNs. The reason is that NNs use a 20-year rolling window for hyperparameter tuning, which helps them to better learn nonlinear relationships, and short- and long-term dependencies (in the case of LSTM) from the data. 

\subsection{Timing Strategy}

The main empirical analysis follows the optimal portfolio construction employed in \cite{johannes2014sequential}, wherein the risk aversion parameter 
$\gamma$
is constant and equal to 4. However, recent literature, including works such as \cite{bekaert2019time} and \cite{bauer2023risk}, shows that the risk aversion parameter might be time-varying. We mitigate the potential issues arising from the selection of a constant
$\gamma$ by implementing the timing strategy in \cite{kelly2022virtue}.\footnote{We thank the anonymous referee for suggesting this robustness check.} Since the timing strategy employs predicted returns as a weight of the market returns, it offers a more straightforward method to assess the predictability of various models. Furthermore, implementing the timing strategy allows us to compare the impact of deep learning models on optimal portfolios with the results documented by \cite{kelly2022virtue}.

Using the previous notations, let 
$r_{t+1}$
and 
$\hat{r}_{t+1}^{\mathcal{M}_s}$
denote the observed and predicted excess log return, where
$\mathcal{M}_s$
is a particular framework used to forecast the excess stock market returns. For each model $\mathcal{M}_s$ (either a linear model or a machine learning approach), 
the prediction 
$\hat{r}_{t+1}^{\mathcal{M}_s}$
is based on a set of variables observed until $t$ as described in Section \ref{section: Evaluating Predictability via Portfolio Performance}. The return of the timing strategy is defined as 
$r_{p,t+1} = \hat{r}_{t+1}^{\mathcal{M}_s} \cdot r_{t+1}.$ For better comparability with the main empirical analysis, we update
$\hat{r}_{t+1}^{\mathcal{M}_s}$
quarterly (annually) and use the same value as a portfolio position within a quarter (year). 

\begin{table}[t!]
\centering
\footnotesize
\caption{Timing strategy}
\begin{minipage}{\textwidth} 
This table reports monthly Sharpe ratios of a timing strategy for different models outlined in Section \ref{section: optimal portfolios}. The table shows the results when the investor changes the allocations quarterly. The table computes the statistics for the whole sample, with expansion and recession periods as defined by NBER, and each of the seven decades since WWII. The portfolio construction starts in February 1955. The sample period spans from January 1945 to December 2018.
\end{minipage}
\vspace{10pt}

\begin{tabular}{lrrrrrrrrr}
\toprule
& \multicolumn{1}{c}{EH} & \multicolumn{1}{c}{OLS1} & \multicolumn{1}{c}{OLS2} & \multicolumn{1}{c}{OLS3} & \multicolumn{1}{c}{OLS4} & \multicolumn{1}{c}{NN1} & \multicolumn{1}{c}{NN2} & \multicolumn{1}{c}{NN3} & \multicolumn{1}{c}{LSTM} \\
\midrule
1955-2018 & 0.052 & 0.047 & 0.032 & 0.053 & 0.065 & 0.078 & 0.072 & 0.115 & 0.113 \\
[0.5em]
Expansions & 0.048 & 0.075 & 0.022 & 0.021 & 0.069 & 0.067 & 0.058 & 0.113 & 0.105 \\
[0.5em]
Recessions & 0.074 & -0.179 & 0.071 & 0.199 & 0.060 & 0.135 & 0.145 & 0.162 & 0.197 \\
[0.5em]
1955-1959 & 0.219 & 0.177 & 0.185 & -0.130 & 0.214 & 0.303 & 0.352 & 0.269 & 0.336 \\
[0.5em]
1960-1969 & 0.109 & 0.071 & 0.029 & 0.060  & 0.130 & 0.128 & 0.117 & 0.151 & 0.081 \\
[0.5em]
1970-1979 & 0.067 & -0.092 & 0.017 & 0.169 & 0.153 & 0.126 & 0.144 & 0.171 & 0.128 \\
[0.5em]
1980-1989 & -0.030 & -0.018 & 0.003 & 0.063 & -0.034 & 0.061 & 0.022 & 0.077 & 0.143 \\
[0.5em]
1990-1999 & 0.183 & 0.178 & 0.180  & -0.180 & 0.133 & 0.136 & 0.167 & 0.182 & 0.207 \\
[0.5em]
2000-2009 & -0.110 & -0.126 & -0.020 & 0.056 & -0.025 & 0.010  & 0.036 & 0.056 & 0.121 \\
[0.5em]
2010-2018 & 0.035 & 0.217 & -0.070 & 0.256 & 0.204 & 0.164 & 0.107 & 0.117 & 0.066 \\
\bottomrule
\end{tabular}%
\label{tab: timing strategy portfolio returns cer and sr}%
\end{table}%

Table \ref{tab: timing strategy portfolio returns cer and sr} presents the Sharpe ratios for the timing strategy across different predictive models. The key conclusions remain robust: strategies based on deep neural networks (NNs) produce substantially higher Sharpe ratios than linear benchmarks, especially during NBER recessions. For the full sample, NN3 and LSTM achieve the highest Sharpe ratios, although NN3 performs slightly better than the recurrent specification in the most recent decade. These results highlight the instability of the ranking among deep architectures across subperiods, and suggest that the main economic value lies in the use of deep neural networks per se, rather than in any single architecture.

At first glance, this is an intriguing divergence from the extant evidence that "complex" models lead to a long-only strategy as shown by \cite{kelly2022virtue}. However, significant differences between their methods and ours might potentially explain the results. The complexity level of deep machine learning considered in our work is much smaller than "complex" models studied in \cite{kelly2022virtue} where the number of parameters exceeds the number of observations. For instance, they employ a training window of 12 months with hundreds of signals compared to 10 years of monthly observations and only 12 predictor variables in our setting. Thus, machine learning methods considered in our paper are arguably closer to linear models in terms of complexity, and it is unsurprising that they yield short and long positions. Intuitively, the LSTM model leverages its flexibility to better capture time-series patterns and, hence, it better identifies the trends in stock market returns, particularly the price declines. Overall, the evidence presented in our work is not inconsistent with \cite{kelly2022virtue} and strengthens the rationale for modeling expected returns through machine learning by highlighting the benefits of using "complex" machine learning models to capture time-series predictability. 

\section{Conclusion}
\label{section: Conclusion}
This paper evaluates the economic gains of using deep learning methods to construct optimal portfolios. We study the portfolio allocation of a long-horizon investor who uses neural networks to predict future returns when choosing an optimal allocation between a market portfolio and a risk-free asset. We propose and compare various architectures of neural networks, including shallow and deep feedforward NNs, as well as the LSTM specification, which is capable of learning long-term relationships. Three key findings emerge from our investigation. 

First, we show that the robust statistical performance of non-linear machine learning methods, such as neural networks, can lead to substantial and meaningful out-of-sample portfolio gains. These gains are robust across various portfolio performance measures, as well as the inclusion of transaction costs and constraints relating to borrowing and short-selling. Second, we find that using deeper or more complex architectures does not necessarily result in larger economic gains. To identify and benefit from complex nonlinear predictive relationships, investors need to harvest more data. In small samples, shallower networks can be competitive with, or even preferable to, more complex ones. Within the class of neural networks that we consider, deep feedforward architectures and the LSTM specification deliver broadly similar economic performance. The recurrent network often performs particularly well during recessions and for more frequent rebalancing, but does not uniformly dominate simpler designs. Third, we show that neural networks perform well even without additional factors, such as time-varying return volatility, which are often suggested in the literature on linear predictive regressions. Our results demonstrate that deep learning methods can identify these complex features in the data in a non-parametric manner, without any specific modelling assumptions. Furthermore, they show that explicitly modelling time-series dependence provides incremental improvements in the context of static nonlinearities.

Our analysis can be extended in several ways. Examining the interaction between NNs and alternative preference specifications would be interesting. In particular, it is unclear whether an investor with a tail-sensitive utility function or a preference for early resolution of uncertainty could achieve comparable economic gains.  \cite{van2010predictive} present evidence that additional economic restrictions can actually improve the model's performance. Our results indicate that restricting portfolio weights negatively affects the gains of the NNs. It would be interesting to examine whether our evidence holds in settings with additional restrictions, particularly those proposed by \cite{van2010predictive}. Finally, extending our analysis to multiple assets is straightforward and would shed light on the economic significance of forecasting returns across different asset classes using NNs.

\newpage
\onehalfspacing
\bibliography{BiblioPortfolio}
\bibliographystyle{apalike}

\newpage

\clearpage

\end{document}